\documentclass[fleqn,usenatbib]{mnras}

\usepackage{newtxtext,newtxmath}

\usepackage[T1]{fontenc}

\DeclareRobustCommand{\VAN}[3]{#2}
\let\VANthebibliography\thebibliography
\def\thebibliography{\DeclareRobustCommand{\VAN}[3]{##3}\VANthebibliography}

\usepackage{graphicx}
\usepackage{amsmath}

\title[Discovery of the gem system TOI--1752]{A gem system with a lava world and a habitable zone sub-Neptune orbiting TOI--1752}

\author[A.\,Peláez-Torres et al.]{A.\,Peláez-Torres$^{1}$\thanks{\texttt{E-mail: apelaez@iaa.es (APT)}},
F.\,J.\,Pozuelos$^{1}$,
G.\,Morello$^{1,2}$,
M.\,Dévora-Pajares$^{3}$,
K.\,Barkaoui$^{4,5,6}$,
L.\,Gkouvelis$^{1}$,
\newauthor
E.\,Pallé$^{4,7}$, 
K.\,A.\,Collins$^{8}$, 
B.\,V.\,Rackham$^{5,9}$, 
S.\,Geraldía-González$^{4,7}$, 
M.\,Centenera-Merino$^{1}$, 
R.\,Varas$^{1}$, 
\newauthor
E.\,Esparza-Borges$^{4,7}$, 
Z.\,Parlapani$^{10,11,4}$, 
J.\,Flores$^{12}$,
J.\,Aceituno$^{12}$,
P.\,J.\,Amado$^{1}$,
A.\,Burdanov$^{13}$,
\newauthor
Y.\,Calatayud-Borras$^{4,7,14}$, 
D.\,R.\,Ciardi$^{15}$, 
B.-O.\,Demory$^{16}$, 
T.\,Gan$^{17}$, 
S.\,Giacalone$^{18}$, 
M.\,Gillon$^{6}$, 
\newauthor
Y.\,G\'omez~Maqueo~Chew$^{19}$, 
K.\,Kawauchi$^{20}$, 
A.\,Khandelwal$^{19}$, 
J.\,Korth$^{21,22}$,
M.\,Lendl$^{22}$, 
J.\,P.\,de\,Leon$^{23}$, 
\newauthor
J.\,Livingston$^{24,25,26}$, 
N.\,Morales$^{1}$, 
F.\,Murgas$^{4,7}$,
N.\,Narita$^{23,24,4}$, 
J.\,L.\,Ortiz$^{1}$,
H.\,Parviainen$^{4,7}$,
\newauthor
M.\,Pichardo~Marcano$^{19}$,
I.\,Plauchu-Frayn$^{27}$,
D.\,Queloz$^{28}$,
D.\,Rapetti$^{29,30}$, 
J.\,Saito$^{23}$, 
A.\,Sánchez-López$^{1}$,
\newauthor
A.\,B.\,Savel$^{31}$, 
R.\,P.\,Schwarz$^{8}$, 
U.\,Schroffenegger$^{16}$, 
M.\,Serra-Ricart$^{4,7,32}$,
C.\,Stockdale$^{33}$,
A.\,H.\,M.\,J.\,Triaud$^{34}$,
\newauthor
J.\,de\,Wit$^{13}$,
F.\,Zong~Lang$^{16}$
}

\date{Accepted XXX. Received YYY; in original form ZZZ}
 
\pubyear{\the\year{}}

\begin{document}
\label{firstpage}
\pagerange{\pageref{firstpage}--\pageref{lastpage}}
\maketitle

\begin{abstract}
The \textit{Transiting Exoplanet Survey Satellite} (TESS) has delivered a large number of transiting planet candidates around nearby stars by identifying periodic decreases in stellar brightness. Establishing the planetary nature of these signals and determining their fundamental properties is a necessary step toward detailed studies of their internal structure, atmospheres, and formation pathways.
In this work, we investigate the planetary nature of the TOI--1752 system (M1 V, $103.02\pm0.34$\,pc), which hosts two TESS candidates: TOI--1752\,b, a short-period object consistent with a lava-world scenario, and TOI--1752\,c, a sub-Neptune-size planet candidate located in the optimistic habitable zone.
We obtained ground-based multi-color photometric follow-up observations of TOI--1752, which we combined with TESS photometry to assess the nature of both signals. We performed a formal statistical validation using the \texttt{TRICERATOPS} framework, while independently vetting the candidates with the neural-network-based classifier \texttt{WATSON-Net}, which provides a machine-learning assessment of their planetary likelihood based on light-curve morphology, centroid diagnostics, and auxiliary vetting features.
We validate TOI--1752\,b as a bona fide planet with a radius of $1.69\pm0.07\,R_{\oplus}$ and an orbital period of $0.935186^{+0.000001}_{-0.000002}$\,days, and TOI--1752\,c with a radius of $2.29^{+0.13}_{-0.14}\,R_{\oplus}$ and an orbital period of $32.7144\pm0.0004$\,days. The combined analysis confirms TOI--1752 as a new planetary system, places TOI--1752\,c within the optimistic habitable zone of its host star, and identifies TOI--1752\,b as a promising target for atmospheric characterization, with an estimated emission spectroscopy metric (ESM) of up to $\sim8$.
\end{abstract}


\section{Introduction}
\label{sec:introduction}

In recent years, space-based missions such as Kepler \citep{borucki2010kepler} and TESS \citep{ricker2015transiting} have driven a rapid increase in the discovery of small planets, including potentially habitable worlds. This search has enabled detailed characterization of planetary radii, internal structures, atmospheric properties, and formation pathways. The Kepler mission, launched in $2009$, was designed to explore Earth-sized planets in the habitable zone (HZ) around Sun-like stars, aiming to address the occurrence rate of this type of planet. The HZ is the region around a star where liquid water could exist on the surface of a planet with an atmosphere composed of CO$_2$, H$_2$O, and N$_2$ \citep{strughold1953green, shapley1953climatic, huang1959occurrence, kasting1993habitable, underwood2003evolution, selsis2007habitable, kaltenegger2011exploring, kopparapu2013habitable}. The first confirmed Earth-like planet on the HZ of its host star was Kepler--186\,f, which belongs to a multi-planetary system orbiting an M-type star \citep{quintana2014earth}. In fact, the ease of finding small planets orbiting around M dwarfs, driven by their low temperature and size, has made them prime targets for ground-based planet observers. As an example, TRAPPIST-1 is a late M8, Jupiter-sized M dwarf–hosted system with seven small transiting planets, three of which are in the HZ \citep{gillon2017seven}.

The Transiting Exoplanet Survey Satellite (TESS), launched in April of 2018, performs a near-all-sky photometric monitoring of millions of nearby stars looking for temporary drops in brightness caused by planetary transits. TESS utilizes a redder photometric bandpass than Kepler's, enhancing its sensitivity to cooler and smaller stars, such as M dwarfs \citep{sullivan2015transiting, ricker2015transiting, barclay2018revised, ballard2019predicted, gilbert2020first}. To date, TESS has provided the community with 7643 candidates, and confirmed 627 planets. Among these, TOI--700 b, c, and d \citep{gilbert2020first}, TOI--1634 b, and c \citep{luque2022density}, and TOI--270 b, c, and d \citep{gunther2019super} stand as reference multi-planet systems hosting at least one HZ planet.

TESS readily finds close-in hot rocky planets, due to their short orbital periods. If these planets are heated strongly enough by their host stars, their rocky surfaces can melt, forming magma oceans; such planets are known as lava worlds. Enveloped in thin silicate atmospheres, enabled by large day-side temperatures (>\,$850$\,K, assuming Earth crust composition), lava worlds are keystone targets for follow-up characterization with infrared spectroscopy \citep{zilinskas2022observability}. Studies on Kepler super-Earths by \citet{miguel2011compositions} showed that their silicate envelopes were not primarily composed of Na, O$_2$, and SiO, as suggested by models from \citet{schaefer2009chemistry}, but instead exhibited a wide range of compositions. However, silicate atmospheres are known to be tenuous, cloudless, and low pressure, making planets with a dayside lava ocean suitable for emission measurements that permit surface composition characterization \citep{giacalone2022validation}. After being theorized for a while, the first two potential lava worlds were Kepler--10\,b \citep{rouan2011orbital, dumusque2014soap}, and CoRoT--7\,b \citep{leger2009transiting, leger2011extreme, samuel2014constraining}. In recent years, the super-Earth 55\,Cnc\,e has emerged as a benchmark exoplanet, extensively studied and often regarded as the archetypal lava world \citep{dawson2010}.

Contrary to the predictions of classical formation-models, the Solar System is not an archetype but rather one of many possible outcomes of planet formation and evolution mechanisms. In fact, the dominant result of planet formation has no counterpart in our stellar system: worlds with $2$\,--\,$3$\,R$_\oplus$, commonly known as sub-Neptunes. At periods <$100$\,days, stars host an average of $0.5$\,--\,$1$ sub-Neptunes, outnumbering larger planets by a factor of $10$:$1$ \citep{fulton2017california}. Sub-Neptunes have been classified as either dry, rocky-core dominated super-Earths or mini-Neptunes, with H$_2$O and H$_2$ rich envelopes \citep[including temperate "hycean" exoplanets;][]{madhusudhan2021habitability}. Recent evidences also suggest the existence of a third class of sub-Neptunes composed of up to $50$\% of H$_2$O, "water worlds" \citep{bluhm2021ultra, dorn2021hidden, luque2022density, acuna2023interior}.
Although radius and mass determinations are essential to constrain the bulk properties of these planets, they are insufficient to constrain the wide range of theoretical models proposed for sub-Neptune interiors. Atmospheric characterization can break this degeneracy by retrieving atmospheric abundances, constraining the metallicity and mean molecular weight of the atmosphere, and informing about
potential interior compositions \citep{miller2008atmospheric, oberg2011effects, lambrechts2014separating, hsu2023chemodynamical}.

The most favorable planetary scenarios for atmospheric characterization include: ($1$) a lava world that sustains a secondary atmosphere ruled by its magma ocean, if not entirely erased by stellar-driven photoevaporation; ($2$) a puffy sub-Neptune with low density and extended atmosphere; and ($3$) a short period gas giant like TOI-6894\,b \citep{bryant2025transiting}, or WASP--193\,b \citep{barkaoui2024extended}. TIC $287139872$ (TOI--1752) hosts the first scenario and could potentially host the second, making it a key system for atmospheric studies. Only one other known system, the aforementioned TOI--1634\,b and c, shares a similar architecture. Other systems, such as TOI-771\,b, and c \citep{lacedelli2025transiting}, also contain a short-period rocky planet and a longer-period sub-Neptune. However, in these cases, either the inner planet is not sufficiently close-in to sustain a dayside molten surface, or the outer planet does not orbit far enough from the host star to fall within its HZ.

Here we present the discovery and validation of a system of two small planets transiting the nearby ($103.02$\,pc), bright (K\,=\,$10.6$\,mag), M1V dwarf TOI--1752. This system hosts an ultra-short-period (USP) super-Earth (TOI--1752\,b), and an HZ sub-Neptune (TOI--1752\,c). In this study, we describe the TESS observations of the system (Section\,\ref{subsec:observations_tess}), ground-based observations (Section\,\ref{subsec:observations_gbp}), derived precise stellar properties of the host star (Section\,\ref{sec:stellar_charact}), analyze the light-curves and model the planets parameters (Section\,\ref{sec:transit_analysis}), validate the system (Section\,\ref{sec:validation}), show the prospects for mass determination and atmospheric characterization (Section\,\ref{sec:prospects}), and present the comparison between TOI--1752 system and other benchmark planets Section\,\ref{sec:discussion}.

\section{Observations}
\label{sec:observations}

%NOT SURE: In this section, we present all the observations of TOI--1752 obtained with TESS and ground-based follow-up facilities.

\subsection{TESS photometry}
\label{subsec:observations_tess}

\begin{figure}
    \centering
    \includegraphics[width=\columnwidth]{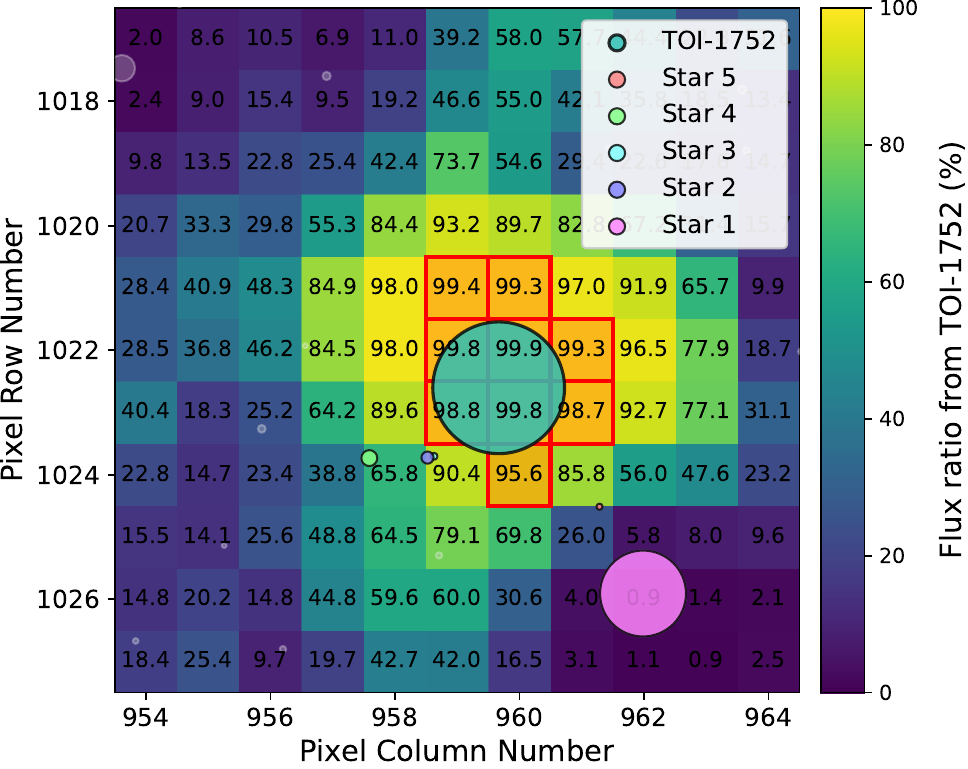}
    \caption{Sector $19$'s contaminating sources within the SPOC aperture (red mask) overlaid on the heat map from TESS observations of the target star, TOI--1752. The sizes of the target and nearby stars are scaled according to their relative fluxes in the TESS filter. The pixel scale is $21\arcsec$ per pixel. This plot was generated using \texttt{TESS-cont} \citep{castro2024toi}.}
    \label{fig:heatmap_s19}
\end{figure}

TOI--1752 has been observed in $42$ TESS sectors ($14$--$26$, $40$--$41$, $47$--$53$, $55$--$60$, and $73$--$86$). Among these, only sectors $19$--$20$, $22$--$23$, $26$, $41$, $47$--$48$, $50$, $55$--$56$, $59$--$60$, $73$, $75$, $78$--$79$, and $82$--$85$ contain transits of TOI--1752.02. In order to reduce the computational cost and discard lower-quality transits, sectors with an RMS\,>\,$6.2$\footnote{\textcolor{blue}{\texttt{RMS value large enough to keep a large sample of good quality sector.}}}
were excluded from further analysis, leaving only sectors $19$ (Nov $28$, $2019$) with short and long cadence\footnote{\textcolor{blue}{\texttt{Short and long cadences are $120$- and $1800$-seconds cadences respectively.}}}, $20$ (Dec $25$, $2019$) with short and long cadences, $22$ (Feb $19$, $2020$) with short and long cadences, $48$ (Jan $28$, $2022$) with short and $600$-seconds cadences, $55$ (Aug $05$, $2022$) with short and $600$-seconds cadences, $56$ (Sep $02$, $2022$) with short and $200$-seconds cadences, $59$ (Nov $26$, $2022$) with short and $200$-seconds cadences, $73$ (Dec $07$, $2023$) with short and $200$-seconds cadences, $82$ (Aug $10$, $2024$) with short and $200$-seconds cadences, and $83$ (Sep $05$, $2024$) with short and $200$-seconds cadences. Transits of TOI--1752.01 and TOI--1752.02 were detected by the Science Processing Operations Center \citep[SPOC,][]{jenkins2016tess} pipeline with and adaptive, noise-compensating matched filter \citep{jenkins2002impact, jenkins2010transiting}, and the image data were reduced and analyzed by the Science Processing Operations Center (SPOC) at NASA Ames Research Center. This produced a threshold-crossing event (TCE) with specific orbital periods for which an initial limb-darkened transit model was fitted \citep{li2019kepler} and a suite of diagnostic tests were conducted to help assess the possible planetary nature of the signal \citep{twicken2018kepler}. A transit search was then carried out with the combined Presearch Data Conditioning Simple Aperture Photometry (PDCSAP; \cite{stumpe2012kepler, stumpe2014multiscale, smith2012kepler}) light curves. The transit signature was also detected for TOI--1752.01 in a search of Full Frame Image (FFI) data by the Quick Look Pipeline (QLP) at MIT \citep{2020RNAAS...4..204H, 2020RNAAS...4..206H}. The TESS Science Office (TSO) reviewed the vetting information and issued an alert on February $27$, $2020$ and February $28$, $2022$, for TOI--1752.01 and TOI--1752.02, respectively \citep{guerrero:TOIs2021ApJS}. The signals were repeatedly recovered by the SPOC pipeline as additional observations were made, including in a recent multi-sector run, $14$-$86$. For the latter, it was found that the host star is located within $1.17\pm2.91$\,\arcsec and $2.94\pm6.00$\,\arcsec of the source of the transit signal for TOI--1752.01 and TOI--1752.02, respectively. For our analysis of the TESS photometric data, we used the \texttt{SHERLOCK} \citep{pozuelos2020gj} pipeline to download the light curves. The TESS image around the position of TOI--1752 in the selected sectors is shown in Fig.\,\ref{fig:heatmap_s19} and Fig.\,\ref{fig:heatmap_sall}.

\subsection{Ground-based photometry}
\label{subsec:observations_gbp}

We used ground-based follow-up photometry from several facilities in the context of the TESS Follow-up Observing Program (TFOP) Sub-Group 1 (SG1) for Seeing-Limited Photometry. The main goals of these observations were to confirm the signals corresponding to TOI--1752.01 and TOI--1752.02 in the target star at transit times, obtain light curves with higher precision than TESS data, and evaluate the transit depths in different bands to assess the chromatic dependence.

Following the approach of \cite{pozuelos2023super}, we have considered for analysis only those observations with a Bayes factor $\Delta\ln Z\,>\,3.2$ in order to preserve the quality of the study (see Table,\ref{tab:observations}), where $\Delta\ln Z\,=\,\ln Z_{\mathrm{transit}}\,-\,\ln Z_{\mathrm{noise}}$. After performing the analysis first with all observations above the Bayes factor threshold and then using only the highest Bayes factor observation per filter, we find that the precision in the parameter estimation of the shorter-period candidate improves when considering only the highest Bayes factor observations for each filter. Therefore, the ground-based observation used to validate TOI--1752.01 is the MuSCAT3 night of June $19$, $2021$.

\begin{figure}
    \centering
    \includegraphics[width=\columnwidth]{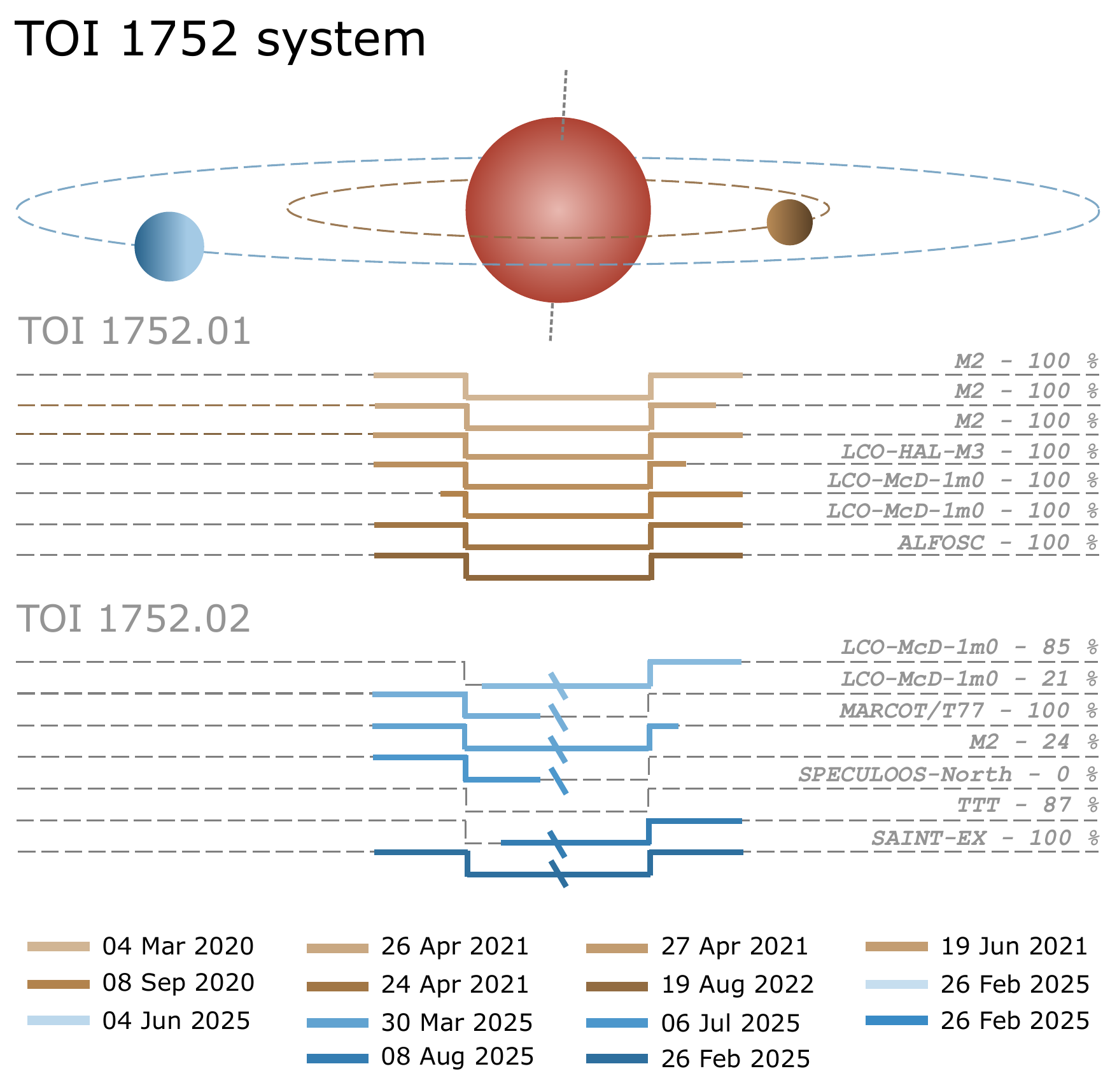}
    \caption{Schematic of ground-based transit photometry observations of TOI--1752 system. The orbital phase coverage of TOI--1752.01 is shown in the top panel (nights of Mar $04$, $2020$, Apr $26$, $2021$, Apr $27$, $2021$, Jun $19$, $2021$, Sep $08$, $2020$, and Apr $21$, $2024$) and that of TOI--1752.02 in the bottom panel (nights of Feb $26$, $2025$, Jun $04$, $2024$, Mar $30$, $2025$, and Jul $06$, $2025$). The instrument/telescope used for the observation and the in-transit coverage percentage are shown in grey. The inclination $i$ is taken into account. 
}
    \label{fig:transits_scheme}
\end{figure}

\begin{table*}
\caption[]{TOI--1752 ground-based observations. All Observations with Bayes factor values $\Delta$\,ln\,Z\,>\,$3.2$ are highlighted with $\dag$. Those observations with the highest Bayes factor per filter are highlighted with $\dag\dag$ (see Section\,\ref{subsec:observations_gbp}).}
\label{tab:observations}
\centering
\scriptsize
\begin{tabular}{llllllll}
  \hline\hline
  \noalign{\smallskip}
  TIC & Observatory & Telescope & Instrument & Filter & Date & Bayes Factor \\
  \noalign{\bigskip}
  \textit{TOI--1752.01} &  &  &  &  &  & \\
  \hline
  \noalign{\smallskip}
  
  287139872.01 & \textit{Teide} & \textit{Carlos\,Sánchez} & MuSCAT2 & \textit{g}, \textit{r}, \textit{i}, \textit{z} & 2020\,Mar\,03 & <\,$3.2$, <\,$3.2$, <\,$3.2$, <\,$3.2$ \\ %<\,$0$, $0.56$, <\,$0$, $0.63$ \\
   &  &  &  & \textit{r}, \textit{i}, \textit{z} & 2021\,Apr\,26 & <\,$3.2$, <\,$3.2$, $3.66^\dag$ \\ %<\,$0$, $2.18$, $3.66^\dag$ \\
   &  &  &  & \textit{r}, \textit{i}, \textit{z} & 2021\,Apr\,27 & <\,$3.2$, <\,$3.2$, <\,$3.2$ \\ %$1.42$, $1.71$, $2.05$ \\
   & \textit{Las\,Cumbres}\,(HAL) & 2-meter & MuSCAT3 & \textit{g}, \textit{r}, \textit{i}, \textit{z} & 2021\,Jun\,19 & $8.91^{\dag\dag}$, $11.61^{\dag\dag}$, $11.76^{\dag\dag}$, $9.15^{\dag\dag}$\\
   & \textit{Las\,Cumbres}\,(McD) & 1-meter & Sinistro & \textit{i} & 2020\,Sep\,08 & $5.11^\dag$ \\
   &  &  &  & \textit{i} & 2021\,Apr\,24 & <\,$3.2$ \\
   & \textit{Roque\,de\,los\,Muchachos} & Nordic\,Optical & ALFOSC & $i$ & 2022\,Aug\,19 & <\,$3.2$ \\
  \noalign{\bigskip}
  \textit{TOI--1752.02} &  &  &  &  & \\
  \hline
  \noalign{\smallskip}
  287139872.02 & \textit{Las\,Cumbres}\,(McD) & 1-meter & Sinistro & \textit{i} & 2025\,Feb\,26 & <\,$3.2$ \\
  &  &  &  & \textit{i} & 2025\,Jun\,04 & <\,$3.2$ \\
  & \textit{Calar\,Alto} & MARCOT-Pathfinder & ZWO ASI1600GT CMOS & R & 2025\,Mar\,30 & <\,$3.2$ \\
  & \textit{La\,Hita} & T77 & QHY174-GPS CMOS & None & 2025\,Mar\,30 & <\,$3.2$ \\
  & \textit{Teide} & \textit{Carlos\,Sánchez} & MuSCAT2 & \textit{g}, \textit{r}, \textit{i}, \textit{z} & 2025\,Jul\,06 & <\,$3.2$, <\,$3.2$, <\,$3.2$, <\,$3.2$ \\
  &  & SPECULOOS-North & Andor iKon-L BEX2-DD & \textit{z} & 2025\,Feb\,26 & <\,$3.2$ \\
  &  & TTT &  COLORS & \textit{i} & 2025\,Aug\,08 & <\,$3.2$ \\
  & \textit{Sierra de San Pedro Mártir} & SAINT-EX Observatory &  Andor Ikon & \textit{I+z} & 2025\,Feb\,26 &  <\,$3.2$ \\
  \hline
\end{tabular}
\end{table*}

\subsubsection{MuSCAT2 observations}
\label{subsubsec:observations_gbp_muscat2}

We observed three transits of TOI--1752.01 with the Multicolour Simultaneous Camera for studying Atmospheres of Transiting exoplanets \citep[MuSCAT2;][]{narita2019muscat2} mounted on the Telescopio \textit{Carlos Sánchez} (TCS) at the Observatorio del Teide (OT), Spain. This instrument is a multi-colour imager able to perform simultaneous photometry in four photometric bands (Sloan-$g$, -$r$, -$i$, and $z_\mathrm{s}$) using four separate CCDs. Each channel has independent exposure times. Usually, when observing M dwarfs, the g filter's CCD is set to have shorter than $15$\,s exposure times and this channel’s images are normally used to auto-guide the instrument.

The MuSCAT2 data were reduced using a dedicated pipeline \citep{parviainen2020muscat2}, which optimized comparison star selection and aperture sizes for extracting the target’s photometric time series.

\subsubsection{LCO-HAL-2m0/MuSCAT3 observations}
\label{subsubsec:observations_gbp_muscat3}

LCO-HAL-2m0 observed a full transit of TOI--1752.01 on UT June $19$, $2021$. It is located at Haleakala Observatory Maui, Hawaii and equipped with the MuSCAT3 multi-band imager \citep{Narita_2020SPIE11447E}.
The observations were conducted simultaneously in the Sloan-$g'$, -$r'$, -$i'$, and Pan-STARRS-$z$ filters (see Fig.\,\ref{fig:lc_M3}).
Science images calibration and photometric extraction were performed using the LCOGT {\tt BANZAI} pipeline \citep{McCully_2018SPIE10707E} and {\tt AstroImageJ}, respectively.

\subsubsection{LCOGT-1m0 observations}
\label{subsubsec:observations_gbp_mcdonal}

We observed one transit of TOI--1752.01 and two of TOI--1752.02 with the Las Cumbres Observatory Global Telescope \citep[LCOGT;][]{brown2013cumbres} $1.0$\,m network nodes installed in the \textit{McDonald Observatory}. The 1\,m telescopes are equipped with $4096\times4096$ SINISTRO cameras, with an image scale of $0.389\arcsec$ per pixel, resulting in a $26\arcmin\times26\arcmin$ field of view. Image calibration and differential photometric data extraction was performed in the same manner as Section\,\ref{subsubsec:observations_gbp_muscat3}.

\subsubsection{MARCOT Pathfinder observations}
\label{subsubsec:observations_gbp_marcot}

We observed one transit of TOI--1752.02 on UT March 30, 2025, using the Multi-Array of Combined Telescopes (MARCOT) Pathfinder \citep{2022SPIE12182E..0MR} installed at the Calar Alto Observatory. The telescope array consists of seven individual GSO 16" f/8 Ritchey–Chrétien Truss Tubes; during this night, six of them were operational, providing a total effective aperture of $\rm{1\ m}$. Six ZWO ASI1600GT CMOS cameras were used. We obtained 214 frames with exposure times of $\rm{120\ s}$ each, but only 160 were retained due to tracking issues.

Because the six telescopes simultaneously observed the same target, the images were stacked during data reduction. The common field of view was $\rm{10.71' \times 14.33'}$, with a plate scale of $0.48$\,\arcsec per pixel. Differential photometry was performed using {\tt AstroImageJ}. We adopted an aperture radius of $6.72$\,\arcsec, and a sky annulus with inner and outer radius of $13.44$\,\arcsec and $26.4$\,\arcsec, respectively. A total of seven reference stars were used, and the observations were conducted using a Bessel R filter.

\subsubsection{La Hita/T77 observations}
\label{subsubsec:observations_gbp_t77}

We observed one transit of TOI--1752.02 on UT March $30$, $2025$ with the $0.77$\,m f/3 telescope (T77) installed in \textit{La Hita} astronomical complex. A QHY174-GPS CMOS camera was used. This model uses the Sony IMX174 CMOS sensor and it has a built-in GPS receiver. In this model, the time stamps in the headers are directly inserted from the GPS unit. $65$\,s exposures were taken from March $30$, $2025$ 21:21:41.14288 till March $31$, $2025$ 05:12:55.18954. No filter was used to maximize the signal. The field of view was $16.6\times10.4$\,\arcmin, with a plate scale of $0.52$\,\arcsec per pixel. The median seeing during the observations was $2.5$\,\arcsec. Autoguiding was applied to avoid significant image drift due to potential tracking problems.

Median dark frames were subtracted from the images, but no flatfield corrections were applied. \texttt{DAOPHOT} routines implemented in IDL were used to derive the photometry. An aperture radius of $5$\,\arcsec was employed and an annulus at an inner radius of $10$\,\arcsec and outer radius of $12.5$\,\arcsec was used to compute the sky background. Besides the target, $7$ reference stars were also measured and used to derive the relative photometry but three of them were rejected and only $4$ references were kept to build a combined reference star.

\subsubsection{NOT/ALFOSC}
\label{subsubsec:observations_gbp_not}

We observed a transit of TOI--1752.01 in the $i$ band with the Alhambra Faint Object Spectrograph and Camera (ALFOSC) mounted on the $2.56$-m Nordic Optical Telescope (NOT) at the \textit{Roque de los Muchachos Observatory}, La Palma, Spain. The observations were obtained on August $19$, $2022$ under consistently good conditions, with an exposure time of $45$\,s. We reduced the photometry and performed an initial transit fit simultaneously using our transit photometry reduction pipeline, following standard aperture photometry practices \citep{Parviainen2019, Parviainen2015}.

\subsubsection{SPECULOOS-North and SAINT-EX}
\label{subsubsec:observations_gbp_speculoos-north}

SPECULOOS-North \citep[Search for habitable Planets EClipsing ULtra-cOOl Stars,][]{Burdanov2022} is a robotic $1.0$\,m telescope located at Teide observatory since $2018$. It is equipped with a 2K$\times$2K Andor iKon-L 936 BEX2-DD detector with a pixel scale of $0.35\arcsec$ and a FOV of $12\arcmin\times12\arcmin$. It is a twin of the  SPECULOOS-South observatory  \citep{Jehin2018Msngr,Delrez2018,Sebastian_2021AA}  located at ESO-Paranal (Chile), and of the SAINT-EX observatory \citep{demory2020} located at the Sierra de San Pedro M\'artir (M\'exico).

SPECULOOS-North and SAINT-EX observed a predicted ingress of TOI--1752.02 on UT February $26$ $2025$ in the Sloan-$z’$ and $I+z$ filters respectively.
The data reduction and analysis were performed using {\tt Prose} pipeline \citep{prose} for SPECULOOS-North data and {\tt PRINCE} (Photometric Reduction and In-depth Nightly Curve Exploration) pipeline for SAINT-EX \citep{demory2020}.

\subsubsection{Two-meter Twin Telescope (TTT)}
\label{subsubsec:observations_gbp_ttt}

TTT (Two-meter Twin Telescope \footnote{\url{https://ttt.iac.es/}} is located at the Teide Observatory on the island of Tenerife (Canary Islands, Spain). Currently, it includes two $0.80$-m telescopes (TTT1 and TTT2) and a 2.0-m telescope (TTT3).  TTT3 is a $f$/$6$ Ritchey-Chrétien telescope that is currently in its commissioning phase. TOI--1752.01 transit images were obtained with COLORS, a $2$k$\times2$k camera mounted at the Nasmyth $2$ focus, equipped with a back-illuminated $13.5~\mu$m\,pixel$^{-1}$ BEX2-DD CCD sensor, resulting in a field of view of $7.85\arcmin\times7.85\arcmin$ and a plate scale of $0.23\arcsec$~pixel$^{-1}$. The transit was observed on UT August $08$ $2025$ in the SDSS i' filter with an exposure time of $30$\,s.

\subsection{Spectroscopic observations}
\label{subsec:observations_gbs}

\subsubsection{IRTF/SpeX}

We observed TOI--1752 with the SpeX spectrograph \citep{Rayner2003} on the 3.2-m NASA Infrared Telescope Facility (IRTF) on 18 May 2025 (UT) under clear conditions with $0.7''$ seeing.
Using the short-wavelength cross-dispersed (SXD) mode and the $0.3'' \times 15''$ slit ($R{\sim}2000$, 0.80–2.42\,$\mu$m) aligned to the parallactic angle, we obtained six 300-s exposures at an airmass of 1.3, nodding in an ABBA pattern.
Standard SXD flat-field and arc-lamp calibrations were taken, along with six 15-s exposures of the A0\,V standard HD\,172728 ($V{=}5.7$) at a similar airmass.
The data were reduced with the Spextool v4.1 pipeline \citep{Cushing2004}, following the same approach as previous analyses \citep{Barkaoui2024, Barkaoui2025, Ghachoui2024}.
The final spectrum has a median SNR of 121 per resolution element.

\subsection{High-resolution imaging}
\label{subsec:observations_hri}

\subsection{3\,m Shane Telescope}

We conducted observations of TOI--1752 on 2021 March 5 using the ShARCS instrument mounted on the 3\,m Shane Telescope at Lick Observatory \citep{kupke2012,gavel2014,mcgurk2014}. The data were acquired with the adaptive optics system operating in natural guide star mode, with the aim of detecting any unresolved stellar companions near the target. Two observational sequences were collected: one using the $K_s$ filter ($\lambda_0 = 2.150\,\mu$m, $\Delta\lambda = 0.320\,\mu$m) and another using the $J$ filter ($\lambda_0 = 1.238\,\mu$m, $\Delta\lambda = 0.271\,\mu$m). The raw frames were processed with the publicly available \texttt{SImMER}\footnote{\url{https://github.com/arjunsavel/SImMER}} reduction pipeline \citep[see, e.g.,][]{savel2020}. The final reduced images and the corresponding contrast curves are displayed in Fig.~\ref{fig:highres_res}. No close stellar companions were detected within the achieved sensitivity limits.

\subsection{5\,m Palomar Telescope}
Observations of TOI--1752 were obtained on June 21, 2021, using the PHARO near-infrared camera \citep{hayward2001} mounted on the 5\,m Hale Telescope at Palomar Observatory, in conjunction with the P3K adaptive optics system operating in natural guide star mode \citep{dekany2013}. PHARO provides a plate scale of $0.025\arcsec$\,pixel$^{-1}$. The observations followed a standard five-point quincunx dither pattern, and the calibrated images were subsequently aligned and combined into a final mosaic, achieving an effective resolution of approximately $0.1\arcsec$.

\begin{figure}
    \centering
    \includegraphics[width=0.95\linewidth]{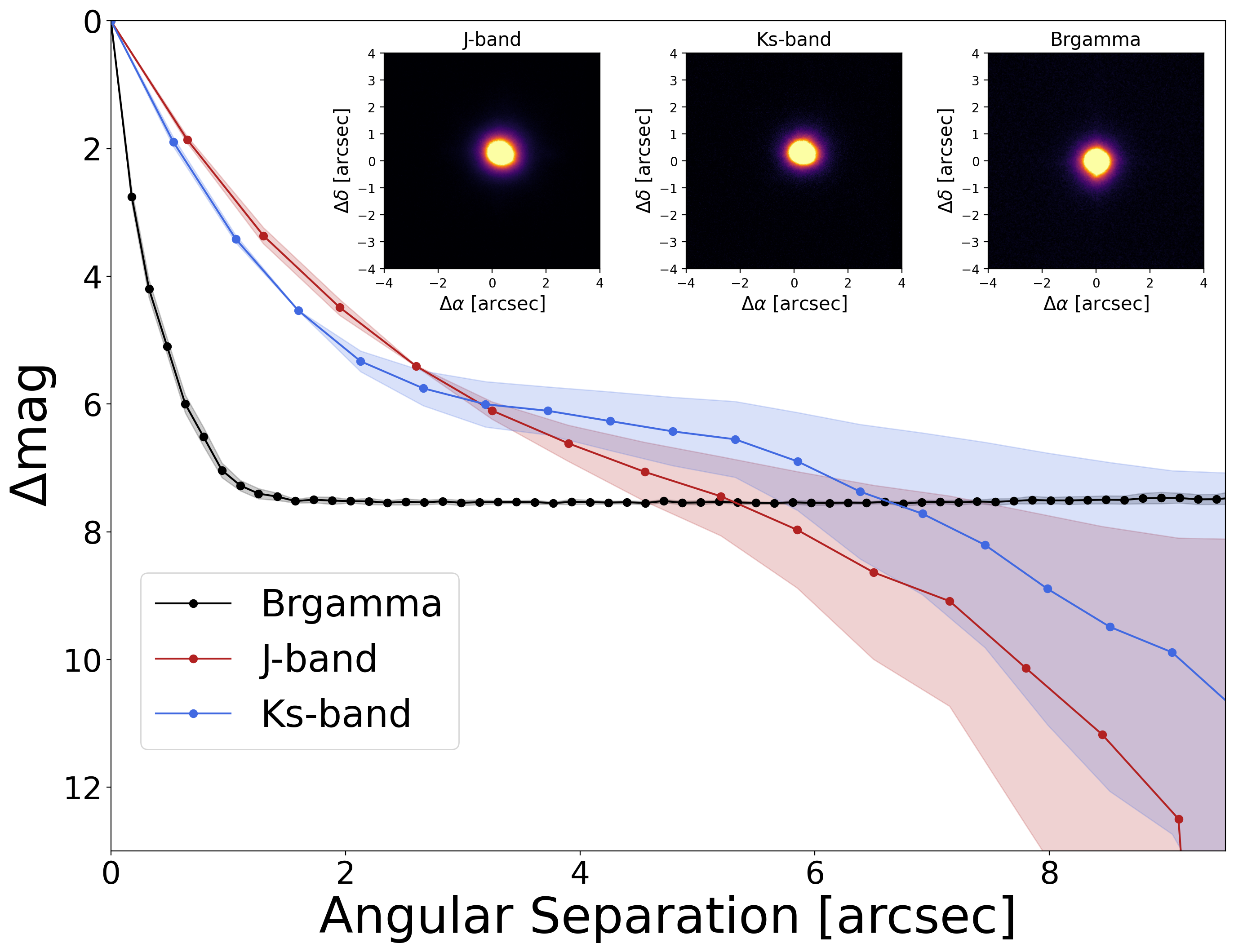}
    \caption{High-resolution imaging and contrast curves as a function of angular separation of TOI--1752 obtained with the PHARO instrument installed at the 5\,m Hale telescope at Palomar Observatory using the Br-$\gamma$ filter and the ShaRCS instrument installed at the Shane telescope using the $K_{s}$ and $J$ filters. No stellar companions were detected within the sensitivity limits.}
    \label{fig:highres_res}
\end{figure}

To estimate the sensitivity of the final adaptive optics images, we carried out a series of artificial companion injection tests. Synthetic sources were placed at $20^\circ$ azimuthal intervals around the target, at separations corresponding to integer multiples of the stellar PSF’s FWHM \citep{furlan2017}. The flux of each injected companion was iteratively adjusted until it reached a 5$\sigma$ detection level based on standard aperture photometry. The resulting contrast curve was derived by averaging the 5$\sigma$ detection thresholds over all azimuthal positions, with the associated uncertainty computed from the RMS scatter of these measurements. No nearby contaminant sources were identified within the detection limits achieved. The final image and contrast curve are presented in Fig.~\ref{fig:highres_res}.

\section{Stellar characterization}
\label{sec:stellar_charact}

\subsection{Spectroscopic analysis}
\label{subsec:stellar_charact_spec_analysis}

\begin{figure}
    \centering
    \includegraphics[width=\columnwidth]{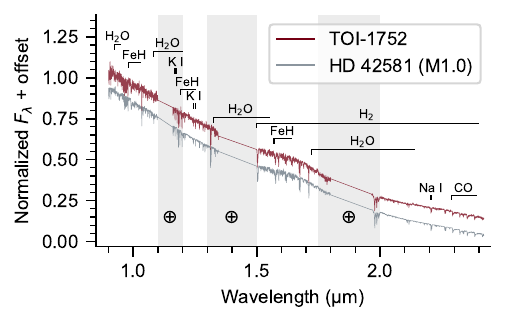}
    \caption{SpeX SXD spectrum of TOI--1752. The target spectrum (red) is shown vertically offset from that of the M1V standard HD\,42581 (grey). Strong spectral features of M dwarfs are indicated, and spectral regions with strong tellurics are shaded.}
    \label{fig:spex}
\end{figure}

The SXD spectrum of TOI--1752 is shown in Fig.\,\ref{fig:spex}.
Comparison with single-star standards from the IRTF Spectral Library \citep{Cushing2005, Rayner2009} using the SpeX Prism Library Analysis Toolkit \citep[SPLAT, ][]{splat} indicates the closest match to the M1\,V standard HD\,42581, and we adopt a near-infrared spectral type of M1.0$\pm$0.5.
From the $K$-band Na\,\textsc{i} and Ca\,\textsc{i} features and the H$_2$O--K2 index \citep{Rojas-Ayala2012}, we estimate $\mathrm{[Fe/H]} = -0.11 \pm 0.11$ using the \citet{Mann2013} relation.

\subsection{Empirical relations}
\label{subsec:stellar_charact_empirical_rel}

We used the empirical relations of \citep{2015ApJ...804...64M, 2019ApJ...871...63M} to derive updated stellar parameters for TOI--1752. 
From its 2MASS $K_s$ magnitude \citep{Skrutskie2006} and Gaia DR3 parallax \citep{GaiaCollaboration2023}, we estimate a stellar mass of $0.524 \pm 0.012\,M_\odot$.
Combining $K_s$ and the SpeX metallicity estimate, we infer a radius of $0.53 \pm 0.015\,R_\odot$.
From the mass and radius, we calculate a surface gravity of $\log g = 4.710 \pm 0.026$\,dex.
The Gaia DR3 BP–RP color yields an effective temperature of $3762 \pm 77$\,K.
Finally, from the radius and effective temperature, the Stefan–Boltzmann Law gives a stellar luminosity of $0.0505 \pm 0.0050\,L_\odot$. Atmospheric and derived stellar parameters are compiled in Table\,\ref{tab:stellar_params}.

\begin{table}
\renewcommand{\arraystretch}{1.3}
\centering
\caption{Calculated and selected parameters for TOI--1752.}
\label{tab:stellar_params}
\begin{tabular}{l c}
\hline
\hline
\noalign{\smallskip}
Parameter & TOI--1752 \\
\noalign{\smallskip}
\hline
\noalign{\smallskip}

\noalign{\smallskip}
\multicolumn{2}{c}{\textit{Stellar atmospheric parameters}} \\
\noalign{\smallskip}
$T_{\rm eff}\,$(K) & $3762\pm77$ \\
$\log{g}\,$(dex) & $4.710\pm0.026$ \\
$[\mathrm{Fe}/\mathrm{H}]\,$(dex) & $-0.11\pm0.11$ \\
Type & M1.0$\pm$0.5 \\
\noalign{\smallskip}
\multicolumn{2}{c}{\textit{Derived stellar parameters} } \\
\noalign{\smallskip}
$R_\star\,(R_{\odot})$ & $0.530\pm 0.015$ \\
$M_\star\,(M_{\odot})$ & $0.524\pm0.012$ \\
$\rho_\star\,(\rho_{\odot})$ & $3.54 \pm 0.31$ \\
\noalign{\smallskip}
\hline
\end{tabular}
\end{table}

\section{Transit analysis}
\label{sec:transit_analysis}

Following the approach by \cite{pozuelos2023super}, we analyzed our light-curves using the \texttt{allesfitter} package \citep{2021ApJS..254...13G}, which allows us to model planetary transits using the \texttt{ellc} package \citep{2016A&A...591A.111M} while also accounting for other phenomena such as stellar flares, spots, and variability. The parameters of interest are retrieved using a Bayesian approach implementing a Markov chain Monte Carlo method using the \texttt{emcee} package \citep{2013PASP..125..306F} or the Nested Sampling inference algorithm \citep[see e.g.,][]{2009MNRAS.398.1601F}. We used the Dynamic Nested Sampling algorithm to directly estimate the Bayesian evidence in this study. This allows us to robustly compare various orbital configurations (see Section\,\ref{subsec:validation_statistical}).

The fitted physical planetary parameters were the ratio of planetary radius over stellar radius ($R_{\rm p}/R_\star$), the sum of the stellar and planetary radius scaled to the orbital semi-major axis (($R_{\rm p}+R_\star)/a$), the cosine of the orbital inclination (cos $i_{\rm p}$), the mid-transit time ($T_{\rm 0}$), and the orbital period ($P_{\rm orb}$). We also added an error-scaling parameter for the white noise as a fitting parameter. To reduce the number of free parameters in the models, we fixed the quadratic limb-darkening (LD) coefficients during the fitting process.

\begin{figure}
    \centering
    \includegraphics[width=\columnwidth]{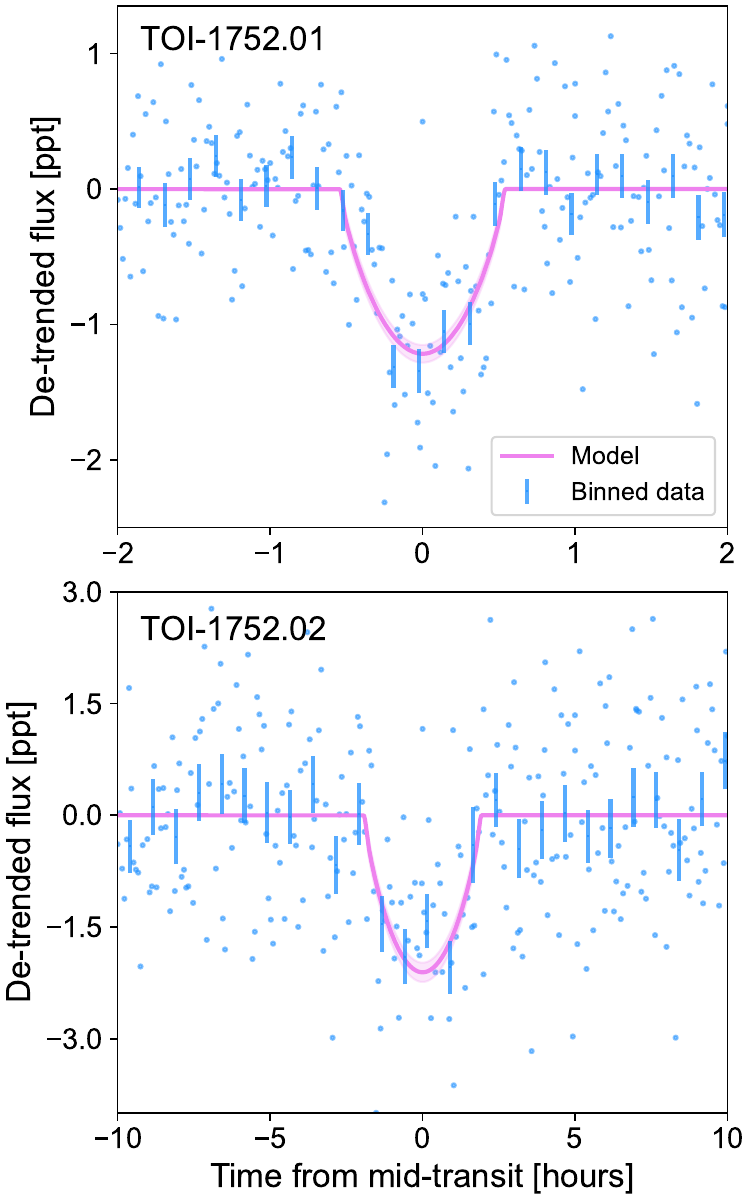}
    \caption{Phase folded light curves from the 10 selected sectors of TESS for planets TOI--1752.01 (top panel) and TOI--1752.02 (bottom panel), together with the transit models (pink line) showing the $1\sigma$ range (pink region) in models during transit. Observed data is binned in steps of $1$ minute (blue dots) and $10$ minutes (blue lines) in phase for TOI--1752.01, and $4.5$ minute (blue dots) and $45$ minutes (blue lines) in phase for TOI--1752.02. Color coding taken from Figure\,$4$ of \protect\cite{gilbert2020first}.}
    \label{fig:lc}
\end{figure}

For each bandpass, we computed the LD coefficients $u_{\rm 1}$ and $u_{\rm 2}$ using \texttt{ExoTETHyS} \citep{Morello2020joss,Morello2020} in combination with \texttt{PHOENIX-COND} stellar atmosphere models \citep{Husser2013,Claret2018}. Next, we transformed the limb-darkening coefficients $u_1$ and $u_2$ into the sampling parameters $q_1$ and $q_2$ (see Table~\ref{tab:ld_coefficients}). These were subsequently implemented as Gaussian priors within the \texttt{allesfitter} framework \citep{2013MNRAS.435.2152K} during the light-curve fitting process. We followed the procedure described in \cite{gunther2019super} to model our data. This procedure consists of several stages described in \cite{pozuelos2023super}, where before performing a global model accounting for all the available data, we analyzed each light curve independently to estimate the GP hyper-parameters and the white noise parameter. The resulting light curves from our model are shown in Fig.\,\ref{fig:lc}. The model and derived physical parameters are reported in Table\,\ref{tab:parameters}. The planetary masses and stellar radial-velocity semi-amplitudes are predicted using \texttt{SPRIGHT} \citep{Parviainen2024}, adopting the theoretical mass–radius relations of \cite{Zeng2019} for FGK host stars.

\begin{table}
\renewcommand{\arraystretch}{1.3}
\centering
\caption{Limb-darkening coefficient values.}
\label{tab:ld_coefficients}
\begin{tabular}{l c c}
\hline
\hline
\noalign{\smallskip}
Band & Prior ($q_1$, $q_2$) & Posterior ($q_1$, $q_2$) \\
\noalign{\smallskip}
\hline
\noalign{\smallskip}
TESS & $0.298\pm0.026$, $0.261\pm0.063$ & $0.329\pm0.061$, $0.227\pm0.060$ \\
$g$ & $0.579\pm0.041$, $0.339\pm0.089$ & $0.53\pm0.11$, $0.23\pm0.11$ \\
$r$ & $0.329\pm0.021$, $0.261\pm0.062$ & $0.304\pm0.058$, $0.267\pm0.059$ \\
$i$ & $0.522\pm0.074$, $0.316\pm0.035$ & $0.457\pm0.050$, $0.260\pm0.047$ \\
$z$ & $0.246\pm0.022$, $0.234\pm0.076$ & $0.236\pm0.064$, $0.257\pm0.066$ \\
\noalign{\smallskip}
\hline
\end{tabular}
\end{table}

\begin{table}
\renewcommand{\arraystretch}{1.3}
\caption[]{TOI--1752 system parameters.}
\label{tab:parameters}
\centering
\begin{tabular}{llll}
  \hline\hline
  \noalign{\smallskip}
  Parameter & TOI--1752.01 & TOI--1752.02 \\
  \noalign{\smallskip}
  \hline
  \noalign{\smallskip}
  
  \noalign{\smallskip}
  \multicolumn{3}{c}{\textit{Transit model parameters}} \\
  \noalign{\smallskip}

  $R_{\rm p}/R_{\star}$ & $0.02953^{+0.00065}_{-0.00086}$ & $0.0400^{+0.0018}_{-0.0023}$ \\
  $b$ & $0.375^{+0.016}_{-0.032}$ & $0.43^{+0.12}_{-0.17}$ \\
  $P_{\rm orb}$\,(days) & $0.9351861^{+0.0000014}_{-0.0000015}$ & $32.71443^{+0.00041}_{-0.00035}$ \\
  $T_0-2450000$ & $9334.47908^{+0.00084}_{-0.00057}$ & $9358.9287\pm0.0049$ \\
  (BJD) &  & \\

  \noalign{\smallskip}
  \multicolumn{3}{c}{\textit{Derived transit parameters}} \\
  \noalign{\smallskip}
  
  $i$\,(deg) & $86.75^{+0.29}_{-0.13}$ & $89.63^{+0.15}_{-0.11}$ \\
  $T_{\rm 14}$\,(h) & $1.041^{+0.023}_{-0.020}$ & $3.59^{+0.19}_{-0.20}$ \\
  $\rho_{\star}$\,$(\rho_{\odot})$ & $4.06 \pm 0.28$ & $3.47 \pm 0.28$ \\

  \noalign{\smallskip}
  \multicolumn{3}{c}{\textit{Derived planet parameters}} \\
  \noalign{\smallskip}
  
  $R_{\rm p}$\,($R_{\oplus}$) & $1.693\pm0.068$ & $2.29^{+0.13}_{-0.14}$ \\
  $a$\,(au) & $0.01624\pm0.00056$ & $0.1614\pm0.0069$ \\
  $T_{\rm eq}$\,(K) & $1036\pm31$
  & $291\pm17$ \\

  \noalign{\smallskip}
  \multicolumn{3}{c}{\textit{Empirical planet parameters}} \\
  \noalign{\smallskip}

  $M_{\rm p}$ & $4.35_{-1.85}^{+1.50}$ & $5.20_{-1.37}^{+2.17}$ \\
  $K_{\star}$\,(m/s) & $4.40_{-1.87}^{+1.53}$ & $1.57_{-0.42}^{+0.70}$ \\
  TSM & $27_{-12}^{+10}$ & $16_{-5}^{+7}$ \\
  ESM & $7.79\pm0.78$ & $0.1$ \\
  
  \noalign{\smallskip}
  \hline
\end{tabular}
\end{table}

\section{TOI--1752 system validation}
\label{sec:validation}

The analysis of the ground-based observations for TOI--1752.01 shows consistent transit depths across the different bands (see Fig.\,\ref{fig:lc_M3}), indicating no significant wavelength dependence. In other words, no chromaticity is detected in the various filters over the wavelength range covered. Therefore, we proceed with the different levels of validation for the planetary nature of the TOI–1752 system.

\subsection{Statistical validation}
\label{subsec:validation_statistical}

We performed a statistical validation analysis using the \texttt{TRICERATOPS} package \citep[see, e.g.,][]{Giacalone2021,giacalone2022validation}. This approach estimates the relative likelihoods of different astrophysical configurations capable of producing the observed transit-like photometric signal, by combining \textit{TESS} light curves, Gaia astrometric and photometric information, contrast curves from high-resolution imaging, and Galactic stellar population models. The \texttt{TRICERATOPS} framework distinguishes between validated planets, likely candidates, and nearby false positives, relying on two parameters: the false positive probability (FPP) and the Nearby False Positive Probability (NFPP). The FPP quantifies the likelihood that the observed event is not of planetary origin, while the NFPP measures the probability that the transit-like signal originates from a nearby contaminant source. Following the standard classification criteria, systems with FPP < 0.015 and NFPP < 0.001 are considered as validated planets, those with FPP < 0.5 and NFPP < 0.001 are likely planetary candidates, and cases with NFPP > 0.1 are treated as probable nearby false positives. For our analysis, we employed the \textit{TESS} photometry described in Section~\ref{subsec:observations_tess}, together with high-resolution imaging contrast curves from the 5-m Palomar Observatory (Section~\ref{subsec:observations_hri}). For TOI--1752.01, we found FPP=0.0027$\pm$0.0014, and, since this candidate was confirmed with ground-based follow-up photometry, its NFPP is zero. For TOI--1752.02, we found FPP = 0.0034$\pm$0.0003 and NFPP = 0.0$\pm$0.0. This indicates that, even without ground-based confirmation, the candidate has a null NFPP value, suggesting that there is no known nearby source capable of producing the detected signal. Hence, under these circumstances, both candidates are placed in the statistically validated planet category. Hereafter, we refer to TOI--1752.01 and TOI--1752.02 as TOI--1752\,b and TOI--1752\,c, respectively.

\subsection{Neural network-based statistical vetting with \texttt{WATSON-Net}}
\label{subsec:validation_watson}

We assessed the planetary nature of both signals with \texttt{WATSON-Net}, a multibranch convolutional neural network trained on Kepler DR25 and evaluated with $10$-fold cross-validation (CV), with an input pipeline standardized to generalize to TESS \citep{2025arXiv251108768D}. Each model processes, in parallel, (i) a global phase-folded light curve ($300$ bins) and four focused views ($75$ bins each) centered on the primary, the expected secondary, and the odd/even subsets within a ±$3$-duration window; (ii) centroid-shift and optical-ghost time series ($75$ points over the same window) to capture source offset and instrumental artifacts; and (iii) a compact set of numerical vetting features (period, duration, radius, counts of well-covered transits, bootstrap FAP, secondary-event metrics, odd–even depth factor) plus basic stellar parameters, all fused after branch-specific convolutions and a branch-dropout layer for robustness to missing diagnostics. Scores are calibrated and mapped to operating regimes; in particular, the isotonic-calibrated model yields a "likely planet" threshold near $0.646$ for a fixed precision of $99$\%, and a more conservative band for "validated planet" ($\sim0.961$ for a perfect precision), whilst "likely negative" ($\sim0.056$) and "validated negative" ($\sim0.007$) regions are also specified. In any case, all these regions must be used for interpretation together with physical models rather than hard stand-alone dispositioning. In deployment, the ten CV models are kept and their outputs are ensembled together; per target we report the ensemble score mean and the across-fold standard deviation as an uncertainty proxy. Applying this to TOI--1752, \texttt{WATSON-Net} returns $0.90\pm0.31$ for TOI--1752\,b, placing it in the likely planet regime albeit with dispersion that grazes the uncertain band, and $0.30\pm0.38$ for TOI--1752\,c, consistent with an uncertain classification in which the machine learning evidence alone is insufficient for validation or refutation mirroring our broader validation picture for the system.

\subsection{Search for additional signals}
\label{subsec:validation_additional}

We employed the \texttt{SHERLOCK} package \citep{pozuelos2020gj,demory2020} to analyze the publicly available \textit{TESS} data (see Section~\ref{subsec:observations_tess}). Our objective was to independently recover the officially-alerted candidates and identify potential candidates that may have been missed by the SPOC and Quick-Look Pipeline (QLP) pipelines \citep[see, e.g.,][]{geo2024, seluck2025, zuniga2025}. For further details on alternative search strategies, refer to \cite{pozuelos2023super} and \cite{devora2024} for the most recent version of the pipeline and a comprehensive description of its functionalities.
Expanding on this approach, we combined all available sectors and investigated orbital periods ranging from 0.5 to 60 days. This range was selected due to computational feasibility and the rapid decrease of the transit probability at longer orbital periods. We successfully recovered the TOI--1752\,b and TOI--1752\,c candidates during the initial SHERLOCK runs, thereby confirming these alerts independently. No additional signals indicative of planetary candidates were identified.

\section{Prospects}
\label{sec:prospects}

\subsection{Prospects for mass determination}
\label{subsec:prospects_mass}

In this section, we evaluate the system's amenability to determining planetary masses using radial velocity measurements. We conducted a set of comprehensive simulations following the procedure described in \cite{barkaoui2025_toi7166} and \citep{10.1093/mnras/stag183}, assuming that the observations are done with the MAROON-X spectrograph \citep{maroonx}.   

To estimate the RV follow-up requirements, we constructed synthetic radial-velocity datasets by randomly selecting observation epochs within the visibility window and computing the corresponding Keplerian signals for the known planets, using the nominal planetary masses listed in Table~\ref{tab:parameters}. The uncertainty of each simulated measurement was characterized through an effective RV error, $\sigma_{\rm eff}$, derived from the publicly available MAROON-X exposure time calculator\footnote{\url{https://maroon-x-etc.gemini.edu/app}}, which provides signal-to-noise predictions as a function of stellar spectral type and brightness and yields the expected \texttt{SERVAL}-based RV precision for both spectrograph arms. Beyond these internal instrumental uncertainties, we incorporated an additional RV jitter component to represent stellar variability and residual instrumental effects not included in the formal per-epoch errors. Large-scale RV surveys have shown that M dwarfs generally display intrinsic velocity scatter at the level of 2-3\,m\,s$^{-1}$ \citep[e.g.,][]{ruh2024}. Motivated by these empirical constraints, we adopt a constant jitter term of 2.5\,m\,s$^{-1}$, which is combined in quadrature with the MAROON-X internal uncertainties to define the final per-epoch error budget used in our simulations. 

\begin{figure}
    \centering
    \includegraphics[width=\columnwidth]{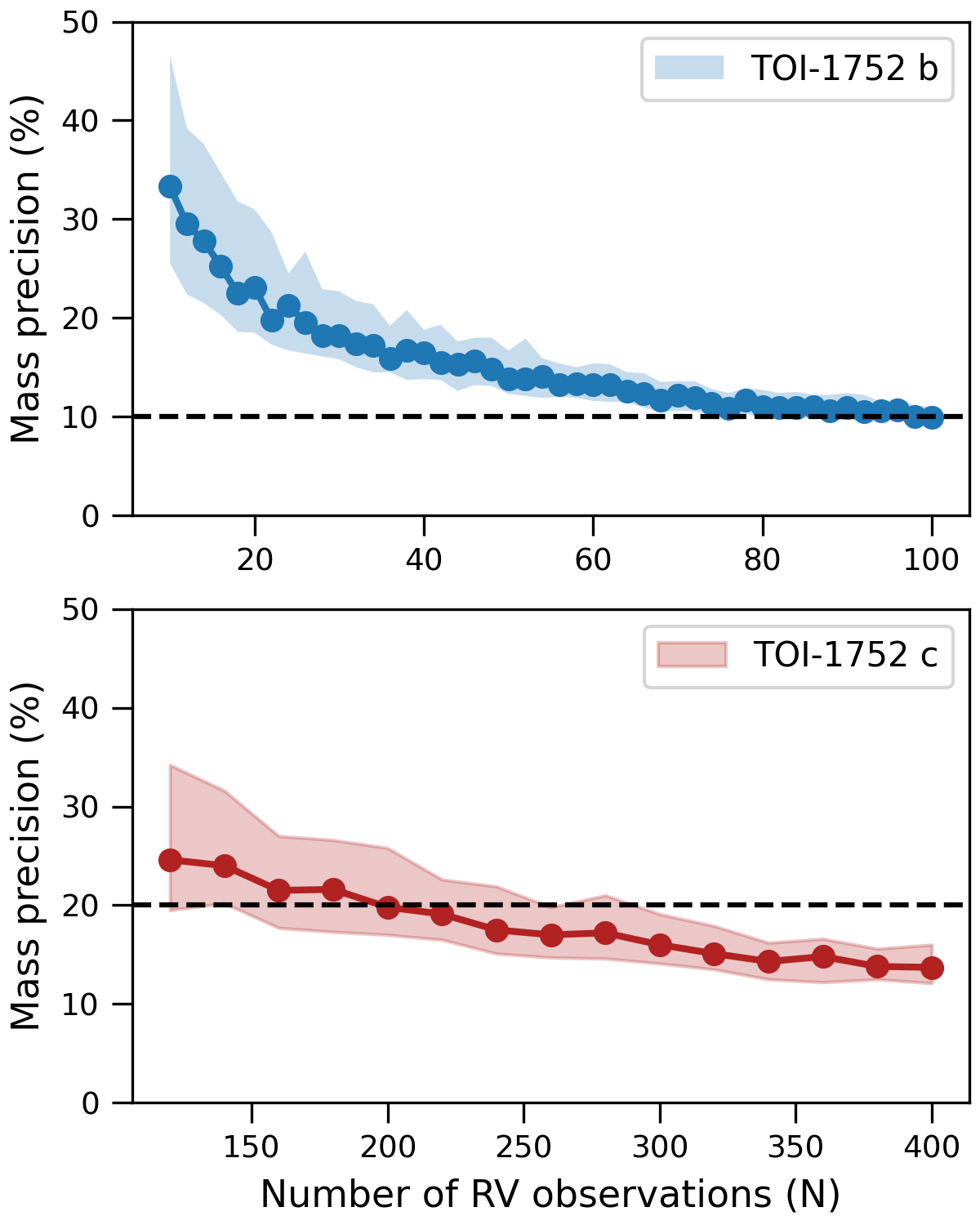}
    \caption{Expected planetary mass precision as a function of the number of radial-velocity (RV) observations for TOI--1752\,b (top panel) and TOI--1752\,c (bottom panel). Solid curves show the median mass precision obtained from Monte Carlo simulations, while the shaded regions indicate the 16$^{th}$–84$^{th}$ percentile range. Horizontal dashed lines mark reference mass precisions of 10\% for TOI--1752\,b and 20\% for TOI--1752\,c.}
    \label{fig:mass-prec}
\end{figure}

We find that the mass of TOI--1752\,b can be measured with a precision of $\sim$10\% using approximately 80 RV observations (corresponding to $\sim$40 hours of on-source time), while achieving a $\sim$20\% precision for TOI--1752\,c (similar to the current precision for K2--18\,b) requires about 200 measurements, roughly 100 hours with MAROON-X (see Fig.~\ref{fig:mass-prec}). Although these observational investments are substantial, they remain within the reach of current high-precision RV facilities and define a realistic pathway toward robust mass determinations for both planets. In this context, TOI--1752 constitutes a benchmark system in which precise planetary masses can be obtained and subsequently combined with atmospheric studies, making it a compelling target for coordinated RV and transmission spectroscopy follow-up (see Section~\ref{subsec:prospects_atm}).

\subsection{Prospects for atmospheric/surface characterisation}
\label{subsec:prospects_atm}

\subsubsection{Surface characterisation and atmospheric degeneracies}
\label{subsubsec:prospects_atm_surface}

The characterisation of USP rocky planets such as TOI--1752\,b offers a unique opportunity to probe the transition between bare-rock and atmosphere-bearing regimes. Recent work has shown that thermal-emission and reflected-light phase curves can directly constrain the physical nature of irradiated surfaces, revealing signatures of heterogeneous lava coverage or cloud-forming atmospheres \citep[e.g.][]{Rouan2011,Hu2015,Demory2016}. A recent \textit{James Webb Space Telescope} (JWST) study of TRAPPIST-1\,b and c demonstrated that combining multi-band phase-curve amplitudes enables the degeneracy between thermal emission from a bare surface and reradiation through a thin atmosphere to be broken \citep{Gillon2025}. Likewise, \citet{Gkouvelis2025} introduced a colour–colour diagnostic framework that links atmospheric opacity and surface emissivity, showing that broadband colours in the near- and mid-infrared can discriminate secondary atmospheres from rocky surfaces, even when clouds or photochemical hazes are present.

For TOI--1752\,b, the equilibrium temperature of $\simeq1000$–$1300$~K lies near the solidus–liquidus transition of common silicate materials. Depending on bulk composition, only part of the dayside may therefore be molten, producing a heterogeneous mixture of lava and solid crust. The fractional coverage of molten terrain can be estimated from thermophysical models that combine the substellar temperature distribution with lithology-dependent melting intervals (e.g. basaltic, feldspathic, peridotitic compositions; \citealt{Leger2011,Kite2016}). Because the thermal emission scales nearly linearly with the molten-surface fraction, future \textit{JWST} or ELT observations could, in principle, invert eclipse or phase-curve data to infer the surface composition. This regime, where the dayside temperature lies precisely at the melting threshold, is particularly diagnostic, since modest variations in composition or albedo can shift the equilibrium between solid and molten phases.  Well-informed forward models that couple realistic optical and photometric properties of planetary analogue materials \citep{Gkouvelis2025b}, and solidus–liquidus phase behavior thus provide a physically grounded method  for retrieving both the surface material and the presence or absence of an atmosphere.

\begin{figure}
    \centering
    \includegraphics[width=\columnwidth]{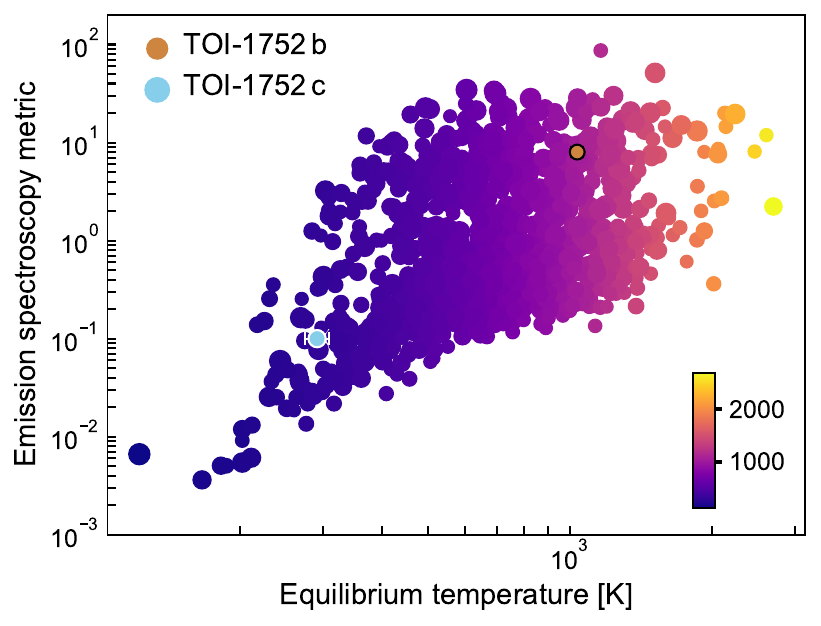}
    \caption{Emission spectroscopy metric as a function of equilibrium temperature of confirmed transiting sub-Neptune exoplanets. TOI--1752\,b and TOI--1752\,c are denoted with a brown and a blue circle, respectively. The different colors correspond to the different $T_{\rm eq}$. Data were extracted from  {\tt NASA Exoplanet Archive.}}
    \label{fig:esm}
\end{figure}

\begin{figure}
    \centering
    \includegraphics[width=\columnwidth]{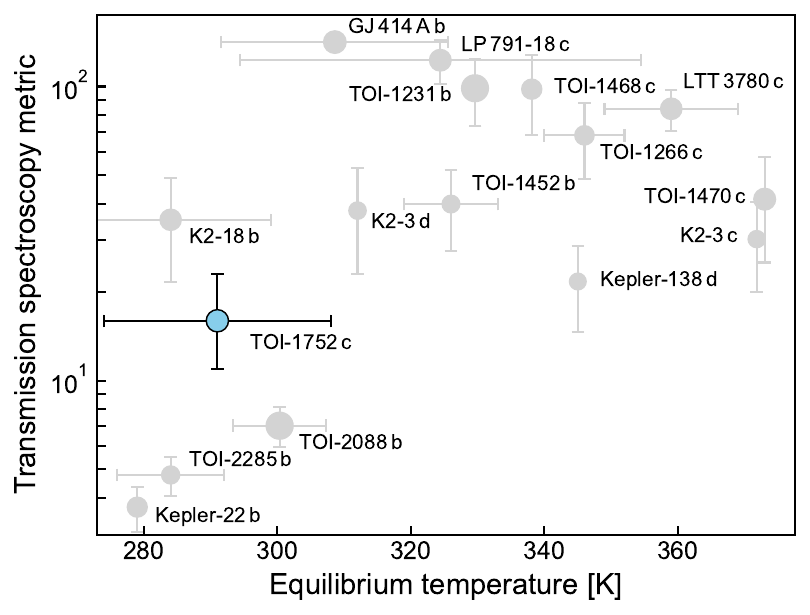}
    \caption{Transmission spectroscopy metric as a function of equilibrium temperature of confirmed transiting temperate sub-Neptune exoplanets. TOI--1752\,c is denoted with a blue circle. Data were extracted from  {\tt NASA Exoplanet Archive.}}
    \label{fig:tsm}
\end{figure}

\begin{figure}
    \centering
    \includegraphics[width=\columnwidth]{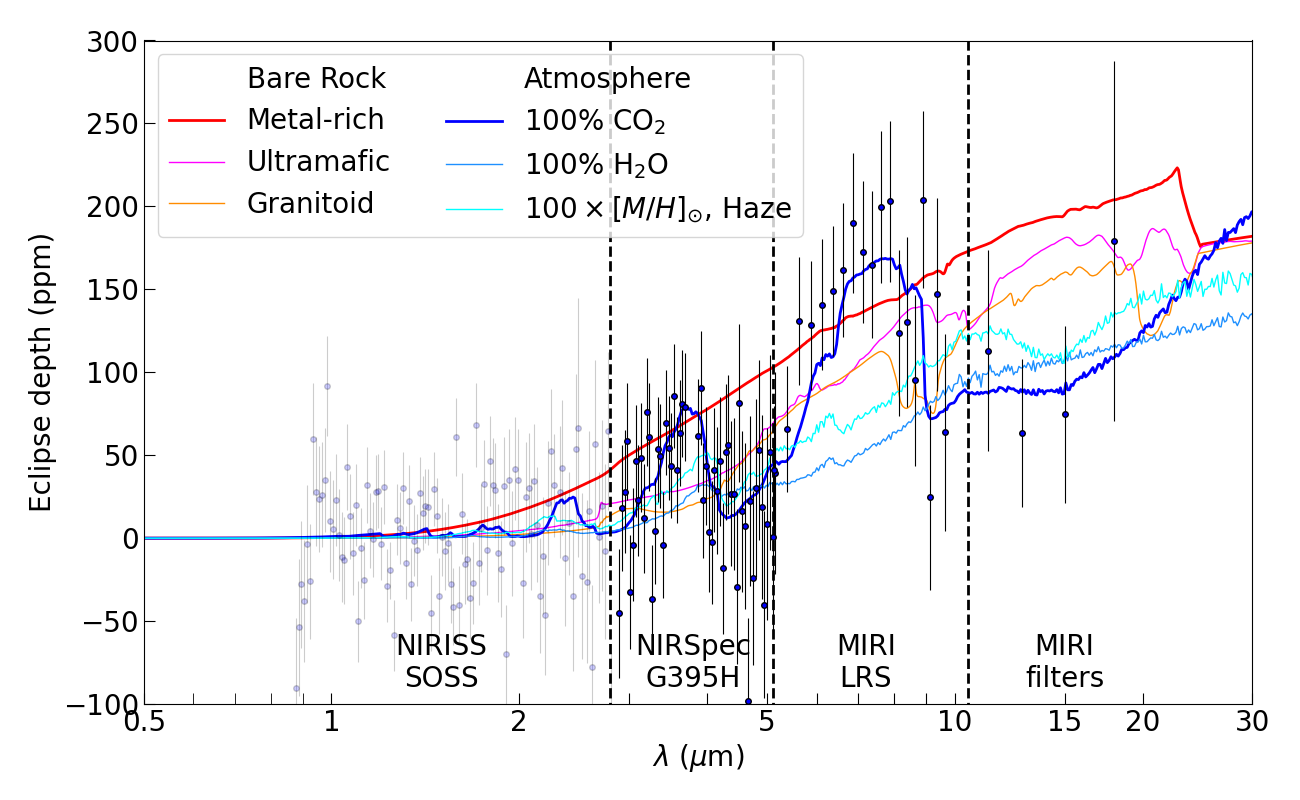}
    \caption{Modelled emission spectra of TOI--1752\,b assuming three atmospheric compositions: 100\,\% CO$_2$ (blue), 100\,\% H$_2$O (light blue), and 100$\times$ solar metallicity with haze (cyan); and three surface compositions: metal-rich (red), ultramafic (magenta), and granitoid (orange).  Points with error bars show the simulated emission spectra obtained from 16 secondary eclipses using the \textit{JWST} NIRISS/SOSS, NIRSpec/G395H, and MIRI/LRS spectroscopic modes, as well as the MIRI photometric filters F1130W, F1280W, F1500W, and F1800W.
    }
    \label{fig:atm_surf_spectra}
\end{figure}

\subsubsection{Potential for \textit{JWST} observations}

Following \cite{Kempton2018}, we evaluated the Transmission Spectroscopy Metric (TSM) and Emission Spectroscopy Metric (ESM) of both validated planets (see Table \ref{tab:parameters}). Figure \ref{fig:esm} shows that TOI--1752\,b lies in the upper part of the sub-Neptune population according to its ESM, making it an attractive target for thermal emission measurements with \textit{JWST}. Figure \ref{fig:tsm} illustrates that TOI‑1752.02 could be the next potentially habitable planet with a detectable atmosphere in transmission, closely resembling the benchmark K2‑18\,b in both equilibrium temperature and TSM \citep[e.g.,][]{madhusudhan2023,madhusudhan2025,Luque2025,Fernandez-Rodriguez2025}.

%IMPORTANT: When estimating the Emission Spectroscopy Metric (ESM), depending on the day-to-nightside radiative transfer assumption, the values range from $\sim4$ up to $\sim8$.

In this work, we generated synthetic spectra and mock observations to assess \textit{JWST}'s capability to detect TOI‑1752\,b's thermal emission and distinguish between plausible atmospheric and surface scenarios.
For the atmospheric models, we considered two plausible secondary atmospheres produced by mantle outgassing, one carbon dioxide-dominated (100\% CO$_2$) and one water-rich (100\% H$_2$O), as well as a less likely primary hydrogen-helium atmosphere with 100$\times$ solar metallicity and haze, included for completeness. The spectral synthesis was performed with \texttt{TauREx} \citep{Waldmann2015a,Waldmann2015b,Al-Refaie2021}, adopting suitable temperature-pressure profiles from the public grid by \cite{Samland2017} and molecular cross sections from the ExoMol database \citep{Polyansky2018,Yurchenko2020,Chubb2021}. 
The bare-rock scenarios considered here include metal-rich, ultramafic, and granitoid compositions, consistent with those modelled by \citet{Gkouvelis2025b} (see also \citealp{Hu2012}). They rely on laboratory-measured reflectance and emissivity spectra from the USGS Digital Spectral Library \citep{Clark2007, Kokaly2017}, while acknowledging the limitations of these libraries when applied to the modelling of rocky exoplanets, as described by \citet{Gkouvelis2025c}.

Based on the six atmospheric and surface scenarios described above, we simulated \textit{JWST} observations using \texttt{ExoTETHyS} \citep{Morello2021}. We considered the NIRISS-SOSS (0.6--2.8~$\mu$m), NIRSpec-G395H (2.88--5.20~$\mu$m), and MIRI-LRS Slitless observing modes, adopting the recommended binning schemes of $R \simeq 100$ for NIRISS and NIRSpec \citep{Carter2024}, and a constant $\Delta \lambda = 0.25~\mu$m for MIRI-LRS \citep{Powell2024}. Additionally, we simulated photometric measurements with the MIRI filters, using the \textit{JWST} Exposure Time Calculator (ETC) to estimate realistic error bars \citep{Pontoppidan2016}.

Figure \ref{fig:atm_surf_spectra} shows the model emission spectra along with synthetic \textit{JWST} data, assuming 16 visits per instrumental mode. Such a high number of visits is often required to characterise rocky exoplanets with \textit{JWST} \citep[e.g.,][]{Lacedelli2025,Turner2025}. Indeed, TOI‑1752\,b's thermal emission spectrum remains below $\sim$200 parts per million (ppm). Among all the scenarios considered, the CO$_2$-dominated atmosphere exhibits the largest deviations from a blackbody spectrum, with spectral features potentially detectable using MIRI-LRS Slitless or NIRSpec-G395H. The spectral region between 10 and 20~$\mu$m shows the largest separation between atmospheric and bare-rock models, and in particular the MIRI F1500W filter at 15~$\mu$m appears best suited to discriminate between these classes of scenarios, consistent with observing strategies outlined in the \textit{JWST} Rocky Worlds DDT program \citep{Redfield2024}. Furthermore, the ultra-short orbital period of TOI--1752\,b makes it a promising candidate for phase-curve observations with the same filter, which would enable a clearer distinction between atmospheric and surface scenarios through measurements of nightside emission \citep{Hammond2025}.

\section{Discussion and conclusions}
\label{sec:discussion}

TOI--1752 is a singular system hosting two planets orbiting a nearby M dwarf. In this section, we aim to place TOI--1752 in context with other planetary systems by evaluating its potential for habitability, its suitability for atmospheric studies, and its prospects for future follow-up characterization.

The sizes of the planets in TOI--1752 fall within the observed gap in the transiting planet radius distribution, known as the radius valley \citep{fulton2017california, van2018asteroseismic, cloutier2020evolution}. The similar radii of both planets are consistent with studies of the Kepler multi-planet population, which suggest that planets within a given system tend to share comparable size and orbital spacing \citep{millholland2017kepler, weiss2018california}. In fact, in the context of small planets, both planets are large enough to be likely that they host atmospheres \citep{weiss2014, rogers2015most}.

The estimated size for TOI--1752\,b lands right in the compositional threshold defined by \cite{rogers2015most}, which places the maximum radius for a planet to have a rocky composition at $1.6\,R_{\oplus}$. Given the extremely short orbital period of TOI--1752\,b, the planet is expected to be highly affected by photoevaporation, whereby the inner planet could not retain its primordial atmosphere due to stellar activity removing it. In this context, TOI--1752\,b is likely to be predominantly rocky, with any existing atmosphere being secondary in origin and subject to continuous renewal. Photoevaporation effects decrease drastically with increasing orbital distance from the host star \citep{lopez2013role}; therefore, TOI--1752\,c would be less affected and could retain a larger fraction of its atmosphere. In fact, other studies such as \cite{lopez2014understanding}, place this compositional boundary at $1.75\,R_{\oplus}$, which further supports the rocky nature of TOI--1752\,b while also suggesting a gas-dominated composition for TOI--1752\,c.

\begin{figure*}
    \centering
    \includegraphics[width=0.85\textwidth]{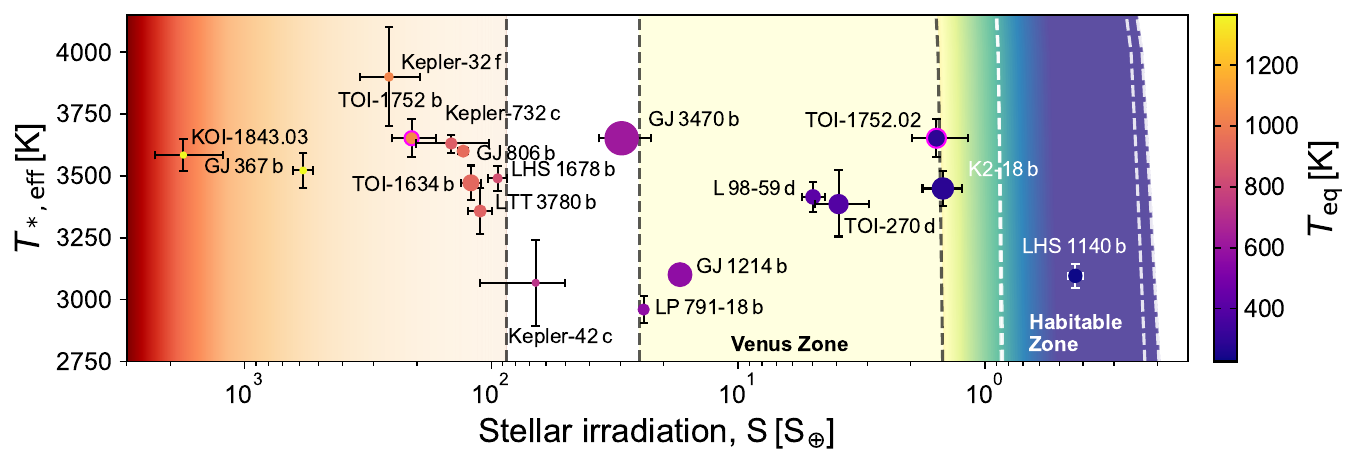}
    \caption{Stellar irradiation comparison between TOI--1752 planetary system (circled in pink), the sub-Neptune sample defined by \protect\cite{madhusudhan2025}, and the sample of USP planets orbiting M dwarfs that are hosted within a multi-planet system, with scaled planetary radii. Planets are colored as a function of the $T_{\rm eq}$. The Venus Zone (VZ) (yellow shade; \protect\citealt{kane2014frequency}) and HZ (blue shade) transition gradually, influenced by complex climate feedback mechanisms that can drive planets toward either Earth-like or Venus-like states within this orbital range \citep{quirino20233d}. The HZ greenhouse variation is represented by the recent Venus, runaway greenhouse, moist greenhouse, maximum greenhouse, and early Mars layers (white dashed lines; \protect\citealt{kopparapu2013habitable}). We also indicate the insolation threshold beyond which the surfaces of planets are expected to melt (the black dashed line on the left; \protect\citealt{zilinskas2022observability}).}
    \label{fig:seff_map}
\end{figure*}

\begin{figure}
    \centering
    \includegraphics[width=\columnwidth]{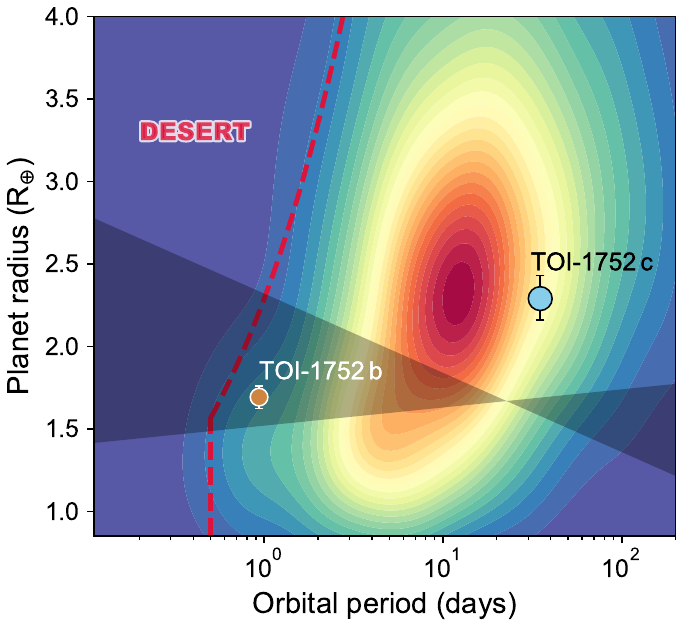}
    \caption{Location of the validated new planets (TOI--1752\,b in brown and TOI--1752\,c in blue) in the period–radius diagram. The different density levels and the corresponding color map indicate planet occurrence, with redder regions being more populated and bluer regions less populated. The red dashed line marks the boundary of the Neptune desert, as defined by \protect\cite{castro2024mapping}. The black-shaded region shows the radius valley of sub-Neptune exoplanets around low mass stars measured by \protect\cite{cloutier2020evolution} and the radius valley for exoplanets around Sun-like stars measured by \protect\cite{martinez2019spectroscopic}, scaled to low mass stars by \protect\cite{cloutier2020evolution}.}
    \label{fig:desert}
\end{figure}

With a radii of $1.693\,R_{\oplus}$ and an orbital period of $0.94$\,days, TOI--1752\,b is one of $31$ known USP planets orbiting M dwarfs, and, similar to other candidates, except for K2--22\,b \citep{sanchis2015}, all have radii smaller than $2\,R_{\oplus}$. TOI--1752\,b is the fourth largest planet among this sample and the third planet with highest known equilibrium temperature $1036$\,K, righ behind Kepler--32\,f \citep{fabrycky2012}. Of the $31$ known USP planets orbiting M dwarfs, at least $10$ have been identified as part of multi-planet systems. These planets are GJ\,367\,b \citep{lam2021}, GJ\,806\,b \citep{palle2023}, LTT\,3780\,b \citep{cloutier2020, nowak2020}, Kepler--42\,c \citep{muirhead2012}, LP\,791--18\,b \citep{crossfield2019}, TOI--1634\,b, LHS\,1678\,b \citep{silverstein2022}, Kepler--732\,c \citep{morton2016}, Kepler--32\,f, and KOI--1843.03 \citep{Ofir2013, rappaport2013} (Fig.\,\ref{fig:seff_map}). Moreover, these systems are aligned, with the outer planets also transiting, consistent with in situ formation scenarios \citep{chatterjee2014}. TOI--1752\,b lies near the boundaries of the demographic arid region in the period–radius diagram known as the Neptune Desert (Fig.\,\ref{fig:desert}). This region has recently been the subject of extensive study, leading many authors to propose updated definitions of its boundaries \citep[e.g.][]{2023A&A...677A..12D, 2023A&A...671A.132S, 2024A&A...692A.162M, 2024A&A...690A..62P}, from the original formulation by \cite{2016A&A...589A..75M} to the more recent and widely adopted definition by \cite{castro2024mapping}.

\begin{figure}
    \centering
    \includegraphics[width=\columnwidth]{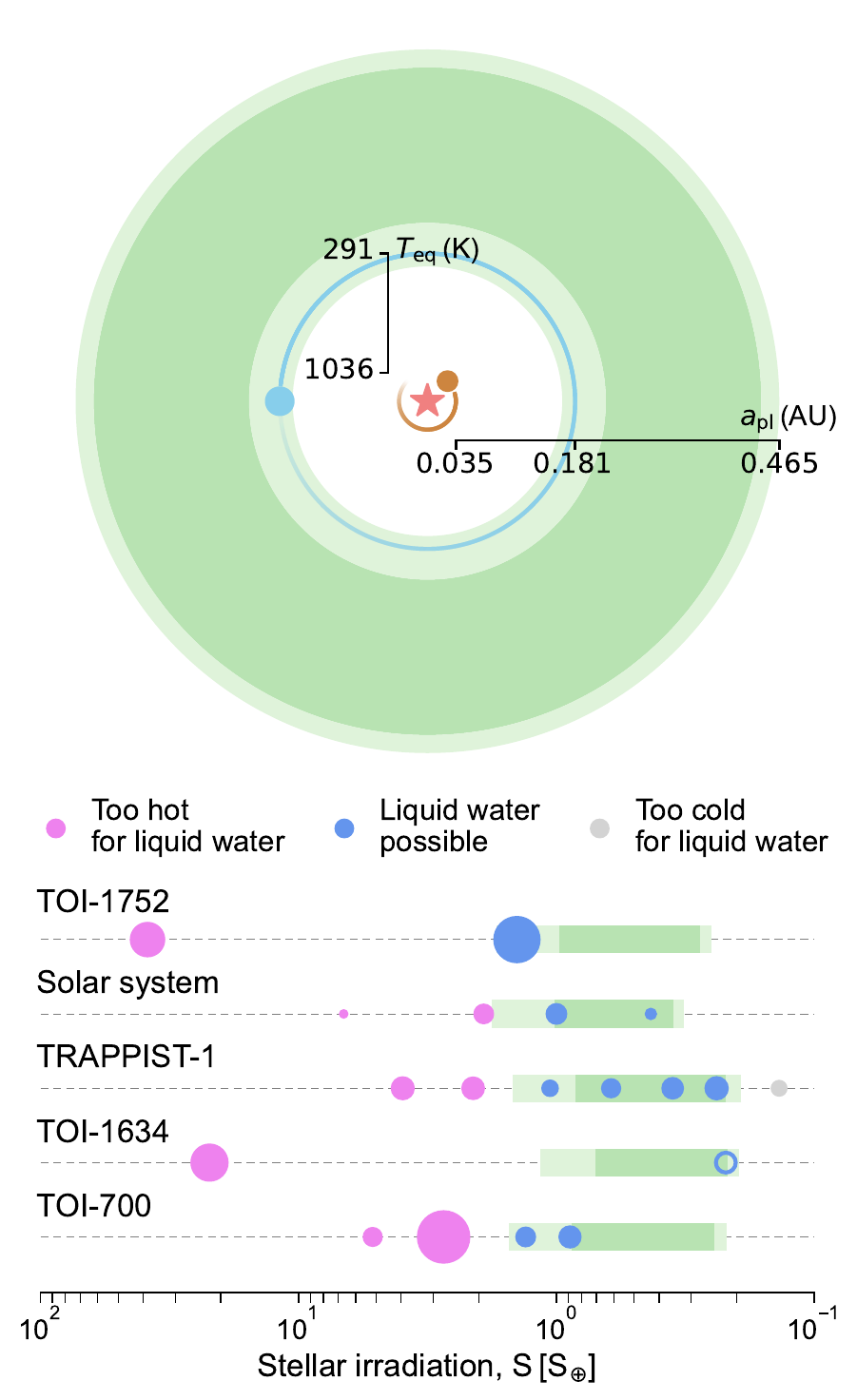}
    \caption{A top-down view of the orbits of the TOI--1752 planets (upper panel). The conservative habitable zone is shown in dark green, and the optimistic habitable zone in light green \citep{kopparapu2013habitable}. We also compare the TOI--1752 system to the solar system and other benchmark exoplanet systems with low-mass host stars and small habitable-zone planets (lower panel). The relative sizes of the planets are to scale. Empty symbols represent non-transiting planets (no radius known).}
    \label{fig:hz}
\end{figure}

With a temperate equilibrium temperature of $291$\,K, a relatively large orbital period of $32$\,days, and a typical sub-Neptunes radius of $2.29\,R_{\oplus}$, TOI--1752\,c can be considered a twin of K2--18\,b. Similarly to K2--18\,b, TOI--1752\,c lands within the habitable zone of its host star (Fig.\,\ref{fig:hz}). Detected almost ten years ago \citep{foreman2015, montet2015}, the super-Earth K2--18\,b orbits an M$2.5$ dwarf with an orbital period of $32.9$\,days, well within the star’s habitable zone. With a radius of $2.38\,R_\oplus$, K2--18\,b was speculated to host carbon-bearing molecules such as CH$_4$ within its H/He envelope, a hypothesis that was later confirmed by \cite{madhusudhan2023} using \textit{JWST} observations. Although K2--18\,b was one of the most if not the most promising candidate for \textit{JWST}, its composition and habitability are still up for debate. Given the similarity in their physical parameters (Fig.\,\ref{fig:seff_map}) and atmospheric properties (Fig.\,\ref{fig:tsm}), we expect TOI--1752\,c, to be as scientifically compelling as K2--18\,b in the upcoming years.

In this study, we have presented and discussed the validation of the planetary nature and the characterization of the TOI--1752 system, including TOI--1752\,b and TOI--1752\,c. Moreover, we have placed TOI--1752 in context by comparing it with other benchmark multi-planet systems from the field, and we have analyzed TOI--1752\,b and TOI--1752\,c as individual planets within the broader population of similar planetary types. In summary, the TOI--1752 planetary system stands out as a promising target for ongoing mass determination efforts and for future comparative atmospheric studies. Notably, TOI--1752\,c, has the potential to stimulate substantial debate in the exoplanetary field, akin to the case of K2--18 b.

\section*{Acknowledgements}

%TESS
Funding for the TESS mission is provided by NASA's Science Mission Directorate.

%SPOC
We acknowledge the use of public TESS data from pipelines at the TESS Science Office and at the TESS Science Processing Operations Center.

Resources supporting this work were provided by the NASA High-End Computing (HEC) Program through the NASA Advanced Supercomputing (NAS) Division at Ames Research Center for the production of the SPOC data products.

%ExoFOP
This research has made use of the Exoplanet Follow-up Observation Program website, which is operated by the California Institute of Technology, under contract with the National Aeronautics and Space Administration under the Exoplanet Exploration Program.

%M2+M3
This article is based on observations made with the MuSCAT2
instrument, developed by ABC, at Telescopio Carlos Sánchez operated on
the island of Tenerife by the IAC in the Spanish Observatorio del
Teide.
This paper is based on observations made with the MuSCAT3 instrument,
developed by the Astrobiology Center and under financial supports by
JSPS KAKENHI (JP18H05439) and JST PRESTO (JPMJPR1775), at Faulkes
Telescope North on Maui, HI, operated by the Las Cumbres Observatory.
This work is partly supported by JSPS KAKENHI Grant Numbers
JP24H00017, JP24K00689, JP21K13955, JP25K17450 and JSPS Bilateral
Program Number JPJSBP120249910.

%NOT
Based on observations made with the Nordic Optical Telescope, owned in collaboration by the University of Turku and Aarhus University, and operated jointly by Aarhus University, the University of Turku and the University of Oslo, representing Denmark, Finland and Norway, the University of Iceland and Stockholm University at the Observatorio del Roque de los Muchachos, La Palma, Spain, of the Instituto de Astrofisica de Canarias. The data presented here were obtained [in part] with ALFOSC, which is provided by the Instituto de Astrofisica de Andalucia (IAA) under a joint agreement with the University of Copenhagen and NOT.

%TTT
This article is based on observations made in the 
Two-meter Twin Telescope (TTT\footnote{\url{http://ttt.iac.es}}) sited at the Teide Observatory of the Instituto 
de Astrofísica de Canarias (IAC), that Light Bridges operates in Tenerife, Canary Islands (Spain). The observation 
time rights (DTO) used for this research were consumed in the PEI "SBSTLLAR25". This research used storage and 
computing capacity in ASTRO POC's EDGE computing center at Tenerife under the form of Indefeasible Computer Rights 
(ICR). The ICR were consumed in the PEI “SBSTLLAR25" with the collaboration of Betchle AG.

%SAINT-EX
Some of the observations presented in this paper were carried out at the Observatorio Astron\'omico Nacional at the Sierra de San Pedro M\'artir (OAN-SPM), Baja California, M\'exico. 
The SAINT-EX team acknowledges support from the Swiss National Science Foundation IZSTZ0\_216537, from the Centre for Space and Habitability (CSH) of the University of Bern, and from the National Centre for Competence in Research PlanetS, supported by the Swiss National Science Foundation (SNSF). 
YGMC, AK, IPF, and MPM are partially supported by UNAM PAPIIT-IG101224.
B.-O. D. acknowledges support from the Swiss State Secretariat for Education, Research and Innovation (SERI) under contract number MB22.00046.

%SNO-Artemis
J.d.W. and MIT gratefully acknowledge financial support from the Heising-Simons Foundation, Dr. and Mrs. Colin Masson and Dr. Peter A. Gilman for Artemis, the first telescope of the SPECULOOS network situated in Tenerife, Spain.
M. G. is FNRS Research Director. He and ULiege thank the Wallonian Region for its contribution to the funding of the SPECULOOS-North/Artemis telescope.

%IAA
IAA-CSIC authors acknowledge financial support from the Severo Ochoa CEX2021-001131-S, PID2022-141216NB-I00 and RYC2022-037854-I grants from MCIN/AEI/ 10.13039/501100011033 and FSE+. In particular, A. P. T. acknowledges the financial support of the FPI-Severo Ochoa contract with reference PRE2022-104692.

%IAC
We acknowledge financial support from the Agencia Estatal de Investigaci\'on of the Ministerio de Ciencia e Innovaci\'on MCIN/AEI/10.13039/501100011033 and the ERDF “A way of making Europe” through projects PID2021-125627OB-C32 and PID2024-158486OB-C32, and from the Centre of Excellence “Severo Ochoa” award to the Instituto de Astrofisica de Canarias.

%BVR
This material is based upon work supported by the National Aeronautics and Space Administration under Agreement No.\ 80NSSC21K0593 for the program ``Alien Earths''. The results reported herein benefited from collaborations and/or information exchange within NASA’s Nexus for Exoplanet System Science (NExSS) research coordination network sponsored by NASA’s Science Mission Directorate.

%Khalid
Funding for KB was provided by the European Union (ERC AdG SUBSTELLAR, GA 101054354).

%Judith
J. K. gratefully acknowledges the support of the Swedish National Space Agency (SNSA; DNR 2020-00104) and of the Swiss National Science Foundation under grant number TMSGI2\_211697.

%Felipe
F. M. acknowledges the financial support from the Agencia Estatal de Investigaci\'{o}n del Ministerio de Ciencia, Innovaci\'{o}n y Universidades (MCIU/AEI) through grant PID2023-152906NA-I00.

%Rapetti
DR was supported by NASA under award number 80NSSC25M7110.

%Hannu
HP acknowledges support by the Spanish Ministry of Science and Innovation with the Ram\'on y Cajal fellowship number RYC2021-031798-I.

%IRTF (Ben)
Visiting Astronomer at the Infrared Telescope Facility, which is operated by the University of Hawaii under contract 80HQTR24DA010 with the National Aeronautics and Space Administration.

%for contamination plot
This work made use of \texttt{TESS-cont} (\url{https://github.com/castro-gzlz/TESS-cont}), which also made use of \texttt{tpfplotter}
\citep{aller2020planetary} and \texttt{TESS-PRF} \citep{bell2022tess_prf}.

\section*{Data Availability}

 %The inclusion of a Data Availability Statement is a requirement for articles published in MNRAS. Data Availability Statements provide a standardised format for readers to understand the availability of data underlying the research results described in the article. The statement may refer to original data generated in the course of the study or to third-party data analysed in the article. The statement should describe and provide means of access, where possible, by linking to the data or providing the required accession numbers for the relevant databases or DOIs.

TESS data used in this work are publicly available from the Mikulski Archive for Space Telescopes (MAST) at \href{https://archive.stsci.edu/}{https://archive.stsci.edu/}.  
Some of the ground-based photometric and spectroscopic data obtained through the TESS Follow-up Observing Program (TFOP) are publicly available or will be made available upon reasonable request to the corresponding author.

\section*{Affiliations}

$^{1}$Instituto de Astrofísica de Andalucía (IAA-CSIC), Gta. de la Astronomía s/n, Genil, E 18008 Granada, Spain\\
$^{2}$INAF- Palermo Astronomical Observatory, Piazza del Parlamento, 1, 90134 Palermo, Italy\\
$^{3}$Dpto. Física Teórica y del Cosmos, Universidad de Granada, 18071 Granada, Spain\\
$^{4}$Instituto de Astrof\'isica de Canarias (IAC), E-38200 La Laguna, Tenerife, Spain\\
$^{5}$Department of Earth, Atmospheric and Planetary Sciences, Massachusetts Institute of Technology, Cambridge, MA 02139, USA\\
$^{6}$Astrobiology Research Unit, Université de Liège, Allée du 6 Août 19C, B-4000 Liège, Belgium\\
$^{7}$Departamento de Astrof\'isica, Universidad de La Laguna (ULL), E-38206 La Laguna, Tenerife, Spain\\
$^{8}$Center for Astrophysics | Harvard \& Smithsonian, 60 Garden Street, Cambridge, MA 02138, USA\\
$^{9}$Kavli Institute for Astrophysics and Space Research, Massachusetts Institute of Technology, Cambridge, MA 02139, USA\\
$^{10}$Denmark Technical University, DK\\
$^{11}$Space Research Organization Netherlands, SRON, NL\\
$^{12}$Centro Astronómico Hispano en Andalucía, Observatorio de Calar Alto, Sierra de los Filabres, 04550, Gérgal, Almería, Spain\\
$^{13}$Department of Earth, Atmospheric and Planetary Science, Massachusetts Institute of Technology, 77 Massachusetts Avenue, Cambridge, MA 02139, USA\\
$^{14}$Departament d'Astronomia i Astrofísica, Universitat de València, 19 Vicent Andrés Estellés Avenue, E-46100 Burjassot, Spain\\
$^{15}$NASA Exoplanet Science Institute - Caltech/IPAC, Pasadena, CA 91125 USA\\
$^{16}$Center for Space and Habitability, University of Bern, Gesellschaftsstrasse 6, 3012 Bern, Switzerland\\
$^{17}$Department of Astronomy, Westlake University, Hangzhou 310030, Zhejiang Province, China\\
$^{18}$Department of Astronomy, California Institute of Technology, Pasadena, CA 91125, USA\\
$^{19}$Universidad Nacional Autónoma de México, Instituto de Astronomía, AP 70-264, Ciudad de México, 04510, México\\
$^{20}$Department of Physical Sciences, Ritsumeikan University, Kusatsu, Shiga 525-8577, Japan\\
$^{21}$Lund Observatory, Division of Astrophysics, Department of Physics, Lund University, Box 118, 22100 Lund, Sweden\\
$^{22}$Observatoire astronomique de l’Université de Genève, Chemin Pegasi 51, 1290 Versoix, Switzerland\\
$^{23}$Komaba Institute for Science, The University of Tokyo, 3-8-1 Komaba, Meguro, Tokyo 153-8902, Japan\\
$^{24}$Astrobiology Center, 2-21-1 Osawa, Mitaka, Tokyo 181-8588, Japan\\
$^{25}$National Astronomical Observatory of Japan, 2-21-1 Osawa, Mitaka, Tokyo 181-8588, Japan\\
$^{26}$Astronomical Science Program, Graduate University for Advanced Studies, SOKENDAI, 2-21-1, Osawa, Mitaka, Tokyo, 181-8588, Japan\\
$^{27}$Universidad Nacional Autónoma de México, Instituto de Astronomía, AP 106, Ensenada 22800, BC, México\\
$^{28}$Cavendish Laboratory, JJ Thomson Avenue, Cambridge, CB3 0HE, UK\\
$^{29}$NASA Ames Research Center, Moffett Field, CA 94035, USA\\
$^{30}$Research Institute for Advanced Computer Science, Universities Space Research Association, Washington, DC 20024, USA\\
$^{31}$Department of Astronomy, University of Maryland, College Park, College Park, MD 20742 USA\\
$^{32}$Light Bridges, SL. Observatorio Astronómico del Teide\\
$^{33}$Hazelwood Observatory, Australia\\
$^{34}$School of Physics \& Astronomy, University of Birmingham, Edgbaston, Birmingham B15 2TT, United Kingdom \\

\bibliographystyle{mnras}
\bibliography{example}

\appendix

\section{All sectors heatmaps}

\begin{figure*}
\centering
\begin{minipage}{0.3\textwidth}
  \centering
  \includegraphics[width=\textwidth]{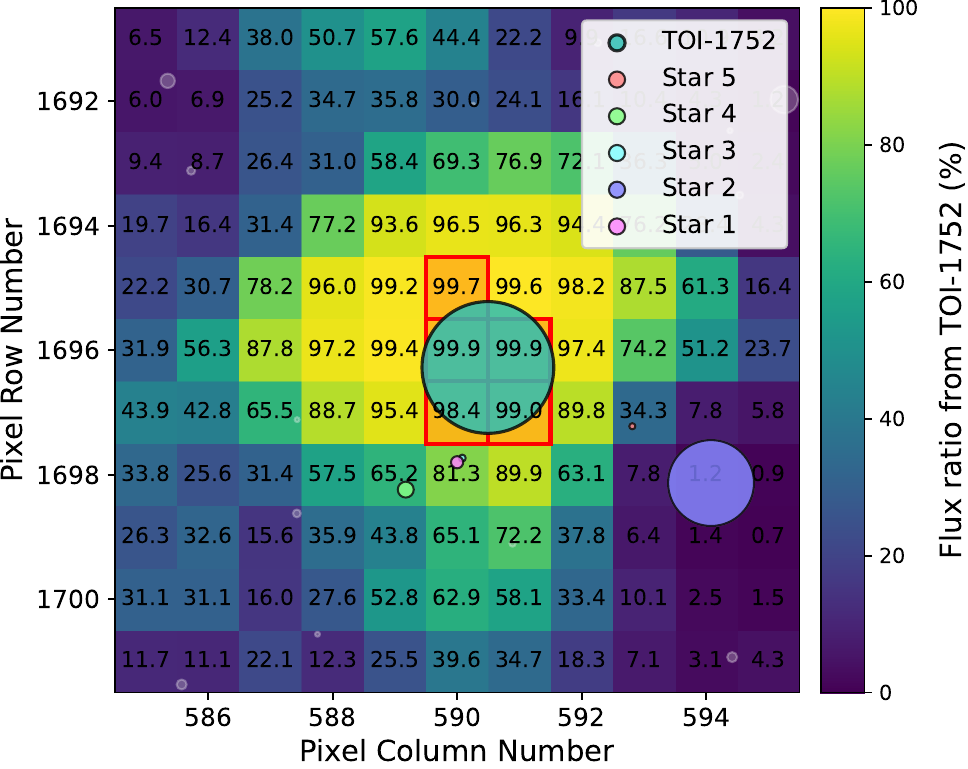}
\end{minipage}
\hfill
\begin{minipage}{0.3\textwidth}
  \centering
  \includegraphics[width=\textwidth]{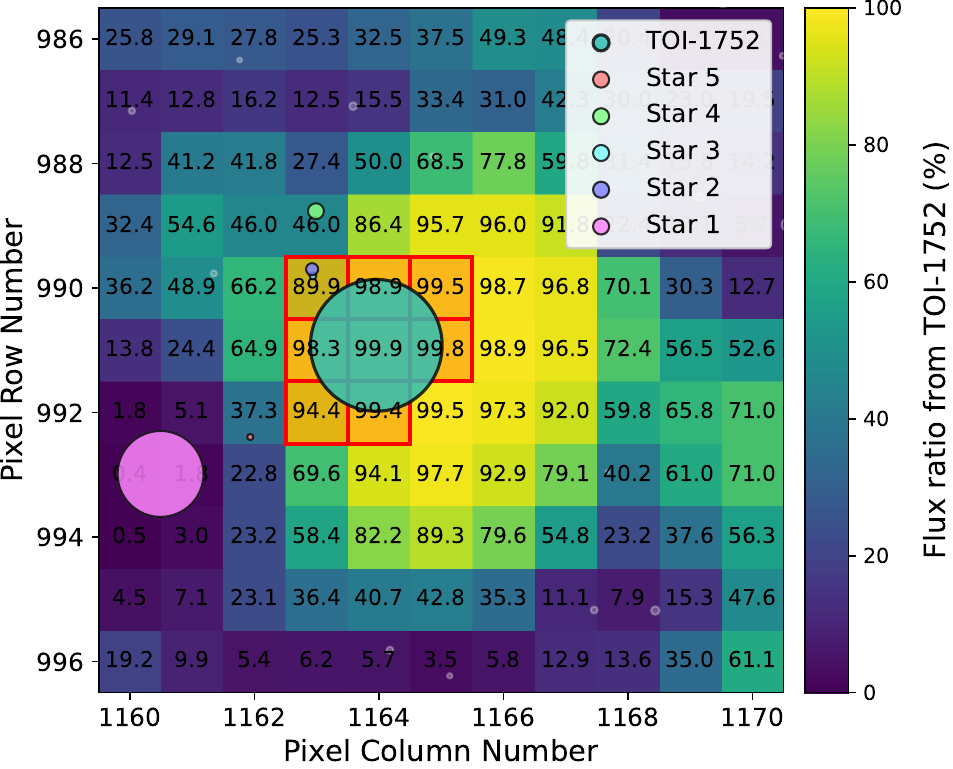}
\end{minipage}
\hfill
\begin{minipage}{0.3\textwidth}
  \centering
  \includegraphics[width=\textwidth]{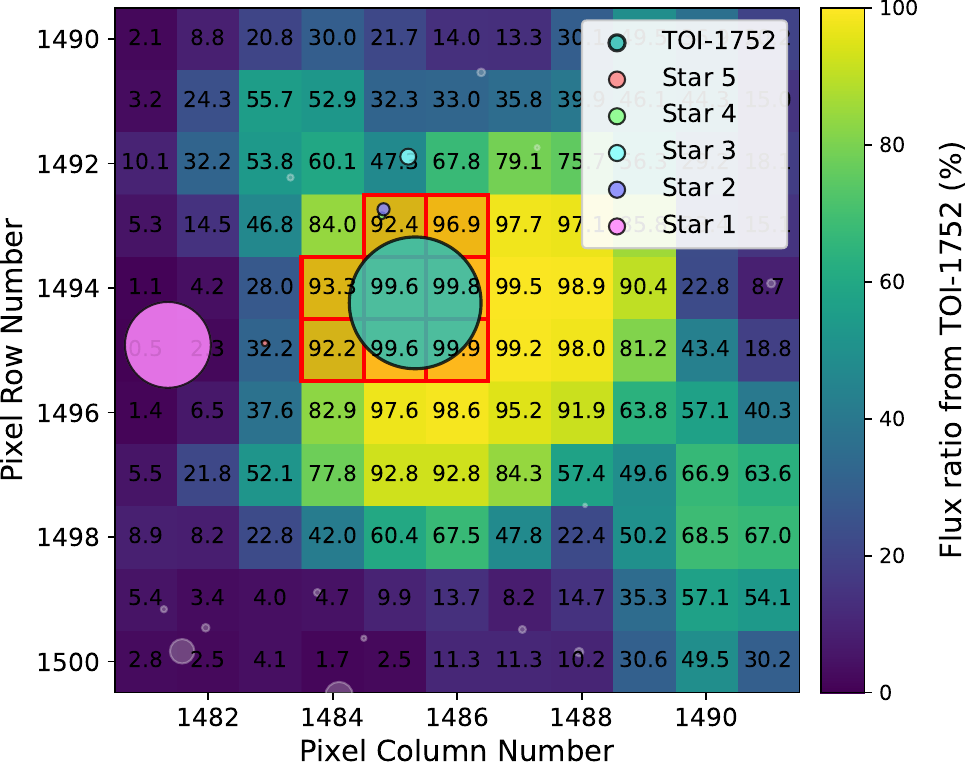}
\end{minipage}

\vspace{0.4cm}

\begin{minipage}{0.3\textwidth}
  \centering
  \includegraphics[width=\textwidth]{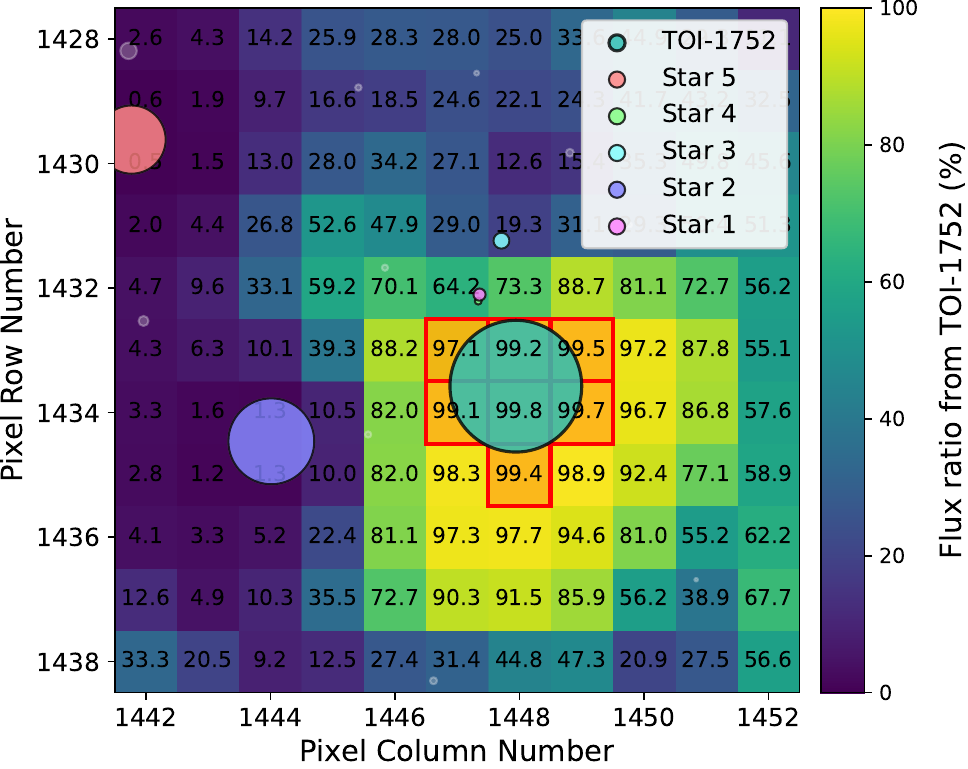}
\end{minipage}
\hfill
\begin{minipage}{0.3\textwidth}
  \centering
  \includegraphics[width=\textwidth]{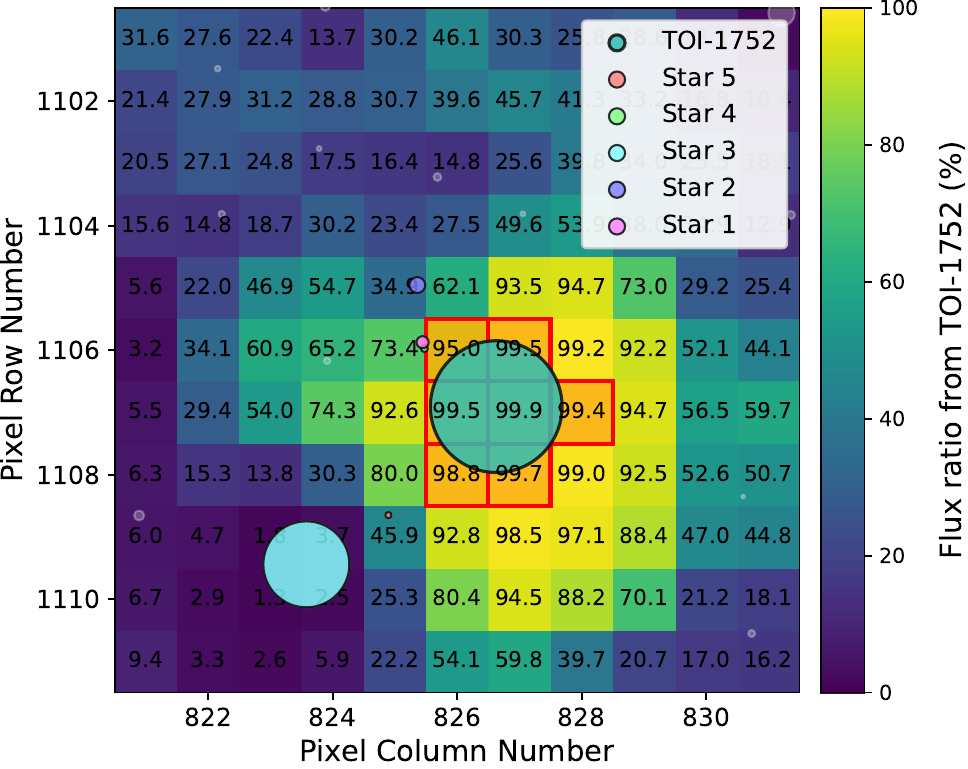}
\end{minipage}
\hfill
\begin{minipage}{0.3\textwidth}
  \centering
  \includegraphics[width=\textwidth]{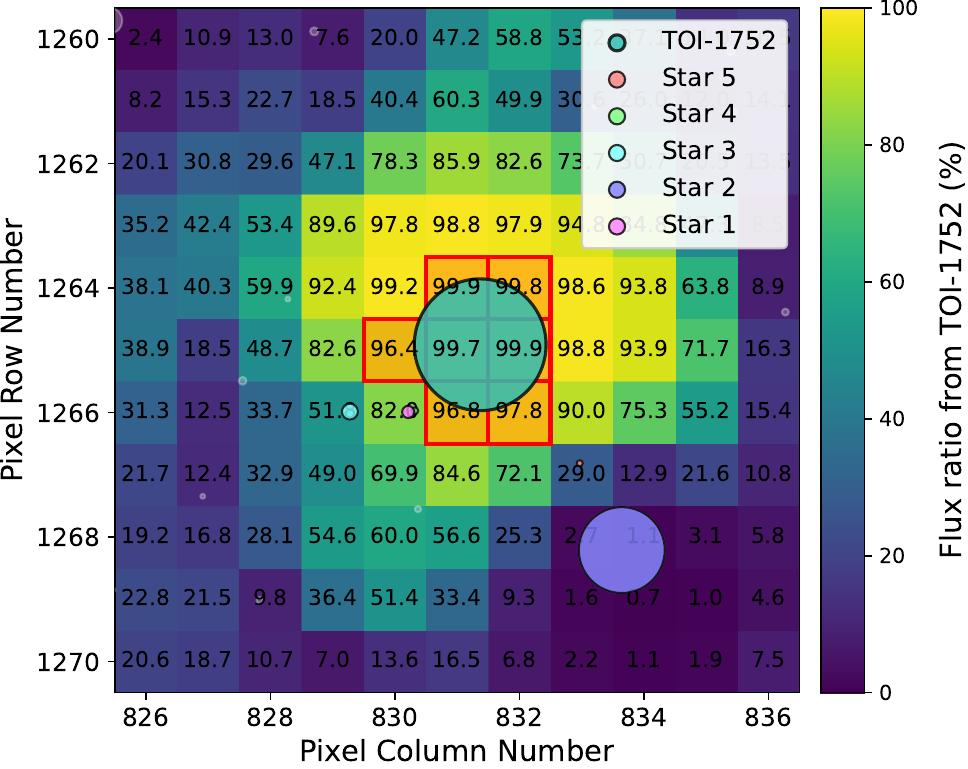}
\end{minipage}

\vspace{0.4cm}

\begin{minipage}{0.3\textwidth}
  \centering
  \includegraphics[width=\textwidth]{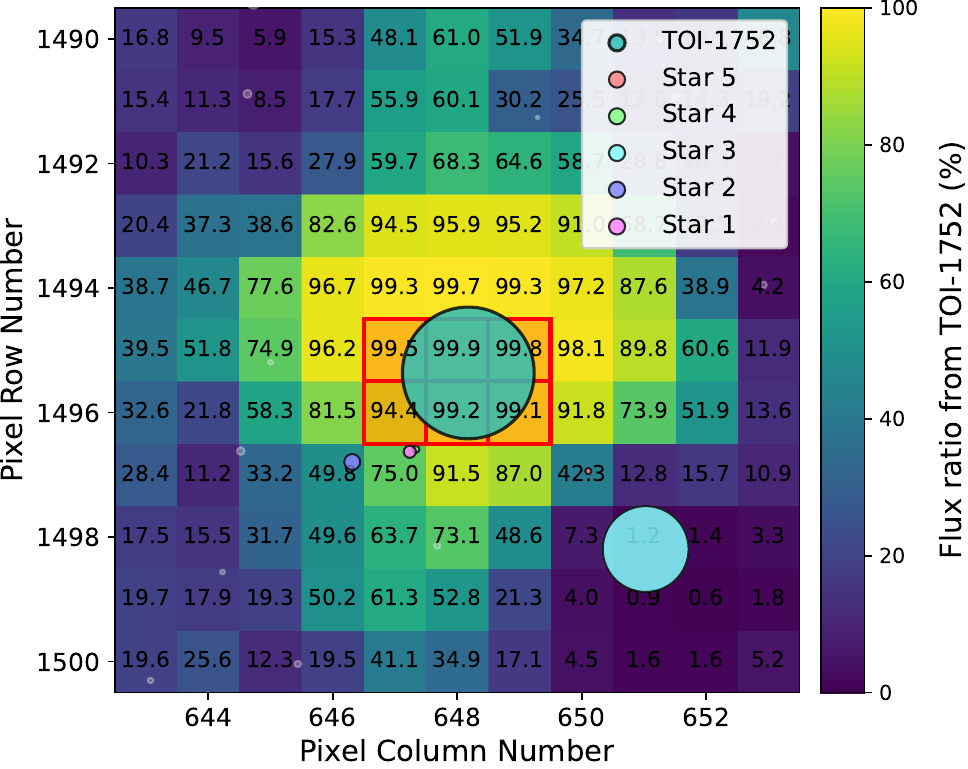}
\end{minipage}
\hfill
\begin{minipage}{0.3\textwidth}
  \centering
  \includegraphics[width=\textwidth]{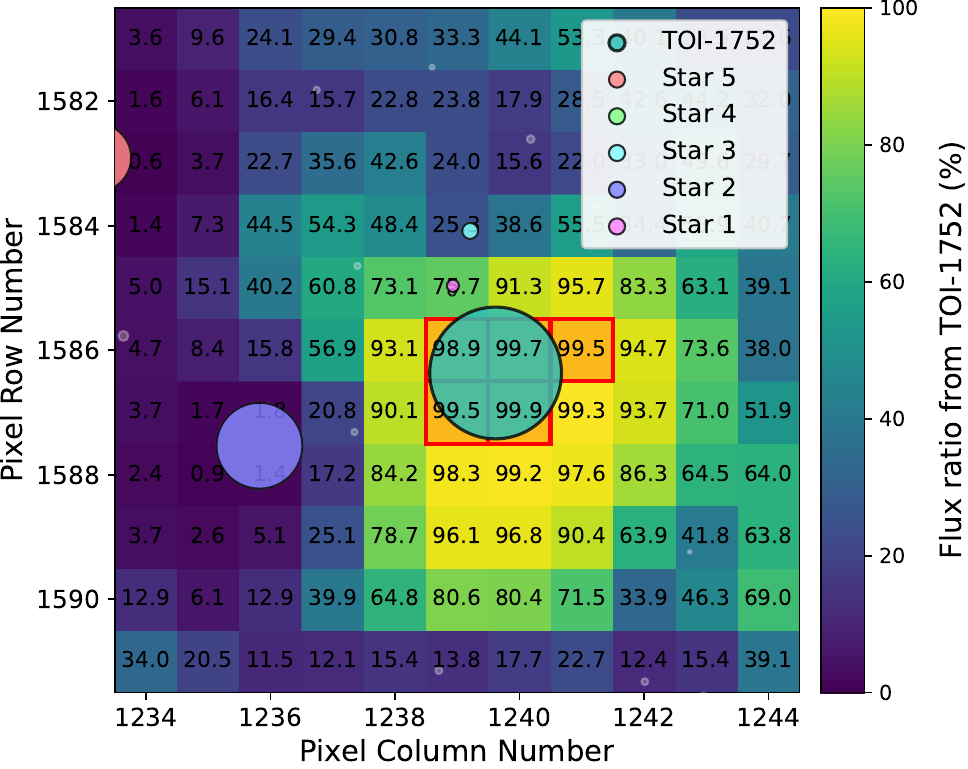}
\end{minipage}
\hfill
\begin{minipage}{0.3\textwidth}
  \centering
  \includegraphics[width=\textwidth]{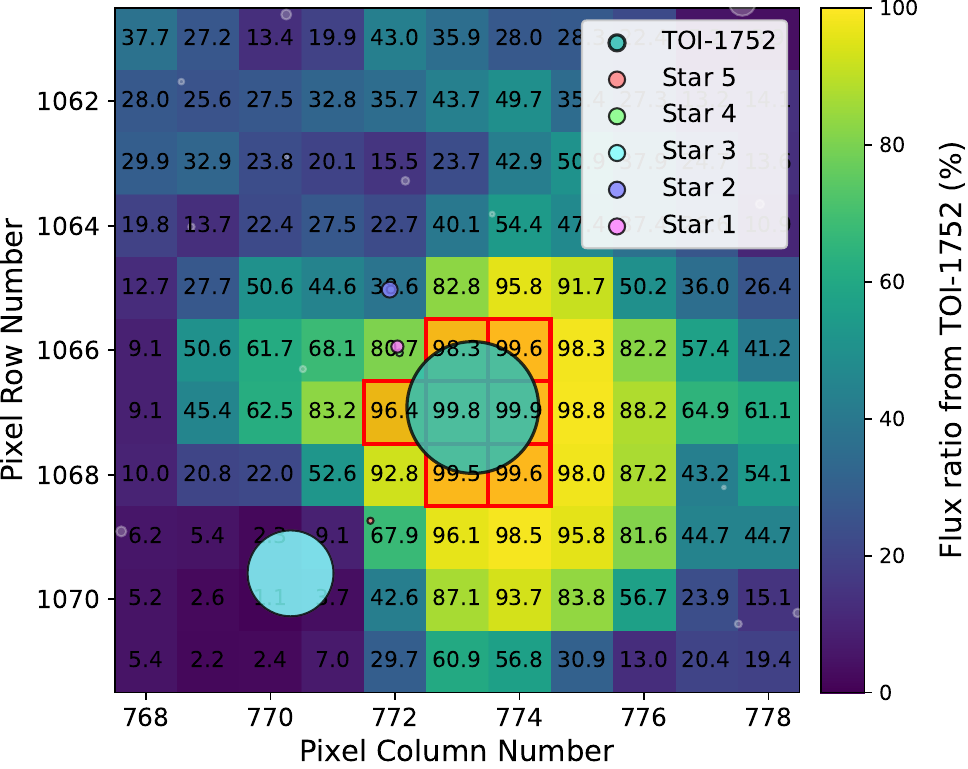}
\end{minipage}

\vspace{0.4cm}

\caption{Same as Fig.\,\ref{fig:heatmap_s19} for sectors $20$, $22$, $48$, $55$, $56$, $59$, $73$, $82$, and $83$ respectively.}
\label{fig:heatmap_sall}
\end{figure*}

\section{MuSCAT3 light curves}

\begin{figure}
    \centering
    \includegraphics[width=\columnwidth]{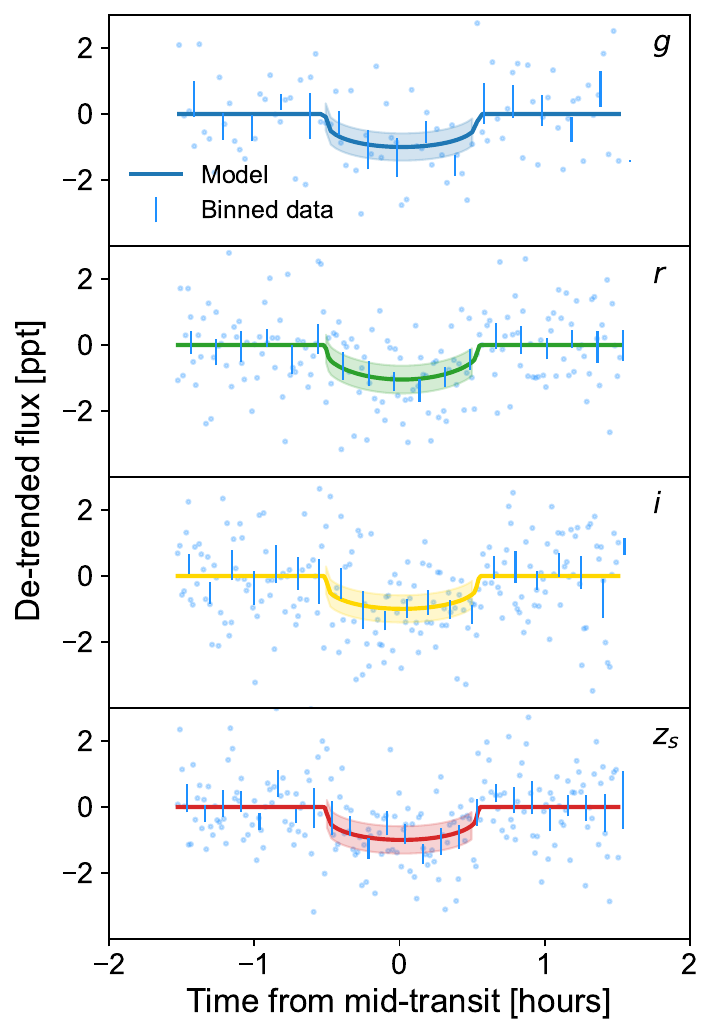}
    \caption{MuSCAT3 light curves for planets TOI--1752.01, together with the transit models (colored line) showing the 1$\sigma$ range (colored region) in models during transit. Observed data is binned in phase in steps of $12$, $10.5$, $9$, and $7.5$ minutes for $g$, $r$, $i$, and $z_\mathrm{s}$ bands respectively.}
    \label{fig:lc_M3}
\end{figure}

\bsp
\label{lastpage}
\end{document}